\documentclass[aoas,MSNbibl,nameyear,seceqn,dvips]{arximspdf}
\usepackage{dcolumn}
\usepackage{graphicx}

\doi{10.1214/13-AOAS627} %kopijuoti is PTS
\volume{7}
\issue{2}
\pubyear{2013}
\firstpage{1162}
\lastpage{1191}

\makeatletter

\newcolumntype{d}[1]{D{.}{.}{#1}}

\newtheorem{theorem}{Theorem}[section]

\newcommand{\cal}{\mathcal}

\newcommand{\R}{{\mathbb R}}
\newcommand{\N}{{\mathbb N}}

\newcommand{\cov}{\operatorname{cov}}
\renewcommand{\hat}{\widehat}

\newcommand{\tr}{\operatorname{tr}}
\newcommand{\diag}{\operatorname{diag}}

\newcommand{\matern}{\mathrm{M}}
\newcommand{\besselk}{\mathrm{K}}

\makeatother

\begin{document}
\begin{frontmatter}

\title{Stochastic approximation of score functions for Gaussian processes\thanksref{T0}}
\runtitle{Approximation of score function}

\thankstext{T0}{\textit{Government License}.
The submitted manuscript has been created by UChicago Argonne, LLC,
Operator of Argonne National Laboratory (``Argonne''). Argonne, a U.S.
Department of Energy Office of Science laboratory, is operated under
Contract No. DE-AC02-06CH11357. The U.S. Government retains for itself,
and others acting on its behalf, a paid-up nonexclusive, irrevocable
worldwide license in said article to reproduce, prepare derivative
works, distribute copies to the public, and perform publicly and
display publicly, by or on behalf of the Government.}

\begin{aug}
\author[A]{\fnms{Michael L.} \snm{Stein}\corref{}\thanksref{t1}\ead[label=e1]{stein@galton.uchicago.edu}},
\author[B]{\fnms{Jie} \snm{Chen}\thanksref{t2}\ead[label=e2]{jiechen@mcs.anl.gov}}
\and
\author[B]{\fnms{Mihai} \snm{Anitescu}\thanksref{t2}\ead[label=e3]{anitescu@mcs.anl.gov}}
\runauthor{M. L. Stein, J. Chen and M. Anitescu}
\affiliation{University of Chicago, Argonne National Laboratory
and\break Argonne National Laboratory}
\address[A]{M. L. Stein\\
Department of Statistics\\
University of Chicago\\
Chicago, Illinois 60637\\
USA\\
\printead{e1}}
\address[B]{J. Chen\\
M. Anitescu\\
Mathematics and Computer\\
\quad Science Division\\
Argonne National Laboratory\\
Argonne, Illinois 60439\\
USA\\
\printead{e2}\\
\hphantom{E-mail: }\printead*{e3}} %adresu isvedimo komanda gale!
\end{aug}

\thankstext{t1}{Supported by the U.S. Department of Energy Grant
DE-SC0002557.}

\thankstext{t2}{Supported by the U.S. Department of Energy through Contract
DE-AC02-06CH11357.}

% HISTORY:
\received{\smonth{5} \syear{2012}}
\revised{\smonth{12} \syear{2012}}

% ABSTRACT
%
\begin{abstract}
We discuss the statistical properties of a recently introduced unbiased
stochastic approximation to the score equations for maximum likelihood
calculation for Gaussian processes. Under certain conditions, including
bounded condition number of the covariance matrix, the approach
achieves $O(n)$ storage and nearly $O(n)$ computational effort per
optimization step, where $n$ is the number of data sites. Here, we
prove that if the condition number of the covariance matrix is bounded,
then the approximate score equations are nearly optimal in a
well-defined sense. Therefore, not only is the approximation efficient
to compute, but it also has comparable statistical properties to the
exact maximum likelihood estimates. We discuss a modification of the
stochastic approximation in which design elements of the stochastic
terms mimic patterns from a $2^n$ factorial design. We prove these
designs are always at least as good as the unstructured design, and we
demonstrate through simulation that they can produce a substantial
improvement over random designs. Our findings are validated by
numerical experiments on simulated data sets of up to 1 million
observations. We apply the approach to fit a space--time model to over
80,000 observations of total column ozone contained in the latitude
band $40^\circ$--$50^\circ$N during April 2012.
\end{abstract}

% KEYWORDS

\end{frontmatter}
\newpage
%s1 #&#
\section{Introduction}\label{sec1}
Gaussian process models are widely used in spatial statistics and machine
learning.
In most applications, the covariance structure of the process is at
least partially unknown and must be estimated from the available data.
Likelihood-based methods, including Bayesian methods, are natural choices
for carrying out the inferences on the unknown covariance structure.
For large data sets, however, calculating the likelihood function
exactly may
be difficult or impossible in many cases.

Assuming we are willing to specify the covariance structure up to some
parameter $\theta\in\Theta\subset\R^p$, the generic problem we are faced
with is computing the loglikelihood for $Z\sim N(0,K(\theta))$ for some
random vector $Z\in\R^n$ and $K$ an $n\times n$ positive definite matrix
indexed by the unknown $\theta$.
In many applications, there would be a mean vector that also depends on
unknown parameters, but since unknown mean parameters
generally cause fewer computational
difficulties, for simplicity we will assume the mean is known to be 0
throughout this work.
For the application to ozone data in Section \ref{sec6}, we avoid modeling the
mean by removing the monthly mean for each pixel.
The simulations in Section \ref{sec5} all first
preprocess the data by taking a discrete Laplacian, which filters
out any mean function that is linear in the coordinates, so that the
results in those sections would be unchanged for such mean functions.
The loglikelihood is then, up to an additive constant, given by
\[
{\cal L}(\theta) = -\tfrac{1}{2}Z'K(\theta)^{-1}Z -
\tfrac{1}{2}\log\det\bigl\{K(\theta)\bigr\}.
\]
If $K$ has no exploitable structure, the standard direct way of calculating
${\cal L}(\theta)$ is to compute the Cholesky decompositon of
$K(\theta)$,
which then allows $Z'K(\theta)^{-1}Z$ and $\log\det\{K(\theta)\}$
to be
computed quickly.
However, the Cholesky decomposition generally requires $O(n^2)$ storage and
$O(n^3)$ computations, either of which can be prohibitive for sufficiently
large $n$.

Therefore, it is worthwhile to develop methods that do not require the
calculation of the Cholesky decomposition or other matrix decompositions
of $K$.
If our goal is just to find the maximum likelihood estimate
(MLE) and the
corresponding Fisher information matrix, we may be able
to avoid the computation
of the log determinants by considering the score equations, which are
obtained by setting the gradient of the loglikelihood equal to 0.
Specifically, defining $K_i =
\frac{\partial}{\partial\theta_i}K(\theta)$, the score equations for
$\theta$ are given by (suppressing the dependence of $K$ on $\theta$)
%
%e1.1 #&#
\begin{equation}\label{score}
\tfrac{1}{2}Z'K^{-1}K_iK^{-1}Z
-\tfrac{1}{2} \operatorname{tr}\bigl(K^{-1}K_i\bigr) = 0
\end{equation}
for $i=1,\ldots,p$.
If these equations have a unique solution for $\theta\in\Theta$,
this solution
will generally be the MLE.

Iterative methods often provide an efficient (in terms of both storage and
computation) way of computing solves in $K$ (expressions of the form $K^{-1}x$
for vectors~$x$) and are based on being
able to multiply arbitrary vectors by $K$ rapidly.
In particular, assuming the elements of $K$ can be calculated as
needed, iterative methods require only $O(n)$ storage, unlike matrix
decompositions such as the Cholesky, which generally require $O(n^2)$ storage.
In terms of computations, two factors drive the speed of iterative methods:
the speed of matrix--vector multiplications and the number of iterations.
Exact matrix--vector multiplication generally requires $O(n^2)$ operations,
but if the data form a partial grid, then it can be done in $O(n\log n)$
operations using circulant embedding and the fast Fourier transform.
For irregular observations, fast multipole approximations can be used
[\citet{anitescu2012mfa}].
The number of iterations required is related to the condition number of
$K$ (the ratio of the largest to smallest singular value),
so that preconditioning [\citet{chenbook}] is often essential; see
\citet{steinchenanitescufiltering} for some circumstances under which
one can prove that preconditioning works well.

Computing the first term in (\ref{score})
requires only one solve in $K$, but the trace
term requires $n$ solves (one for each column of $K_i$) for $i=1,\ldots,p$,
which may be prohibitive in some circumstances.
Recently, \citet{anitescu2012mfa} analyzed and demonstrated a
stochastic approximation of the
trace term based on the Hutchinson trace
estimator [\citet{hutchinson}].
To define it, let $U_1,\ldots,U_N$ be i.i.d. random vectors in $\R^n$
with i.i.d.
symmetric Bernoulli components, that is,
taking on values 1 and $-1$ each with probability $\frac{1}{2}$.
Define a set of estimating equations for $\theta$ by
%
%e1.2 #&#
\begin{equation}\label{ascore}
g_i(\theta,N) = \frac{1}{2}Z'K^{-1}K_iK^{-1}Z
-\frac{1}{2N}\sum_{j=1}^N
U_j'K^{-1}K_iU_j = 0
\end{equation}
for $i=1,\ldots,p$.
Throughout this work, $E_\theta$ means
to take expectations over $Z\sim N(0,K(\theta))$ and over the $U_j$'s as
well.
Since $E_\theta(U_1'K^{-1}K_iU_1) = \operatorname{tr}(K^{-1}K_i)$, $E_\theta
g_i(\theta,N) = 0$ and (\ref{ascore}) provides a set of unbiased estimating
equations for $\theta$.
Therefore, we may hope that a solution to (\ref{ascore}) will provide
a good
approximation to the MLE.
The unbiasedness of the estimating equations
(\ref{ascore}) requires only that the components of the $U_j$'s have mean
0 and variance 1; but, subject to this
constraint, \citet{hutchinson} shows that, assuming the components
of the $U_j$'s are independent, taking them to be symmetric Bernoulli
minimizes the variance of $U_1'MU_1$ for any $n\times n$ matrix $M$.
The Hutchinson trace estimator has also been used to approximate
the GCV (generalized cross-validation) statistic in nonparametric regression
[\citet{girard,wahba}].
In particular, \citet{girard} shows that $N$ does not need to be large
to obtain a randomized GCV that yields results nearly identical to those
obtained using exact GCV.

Suppose for now that it is possible
to take $N$ much smaller than $n$ and obtain an estimate of $\theta$
that is nearly as efficient statistically as the exact MLE.
From here on, assume that any solves in $K$ will be done using
iterative methods.
In this case, the computational
effort to computing (\ref{score}) or (\ref{ascore}) is roughly linear
in the number of solves required
(although see Section \ref{sec4} for methods that make $N$ solves for
a common matrix $K$ somewhat less than $N$ times the effort of one solve),
so that (\ref{ascore}) is much easier
to compute than (\ref{score}) when $N/n$ is small.
An attractive feature of the approximation (\ref{ascore}) is that
if at any point one wants to obtain a better approximation to
the score function, it suffices to consider additional $U_j$'s
in (\ref{ascore}). However, how exactly to do this if using the
dependent sampling scheme for the $U_j$'s in Section \ref{sec4} is not so obvious.

Since this stochastic
approach provides only an approximation to the MLE, one must
compare it with other possible approximations to the MLE.
Many such approaches exist, including spectral methods, low-rank
approximations, covariance
tapering and those based on some form of composite likelihood.
All these methods involve computing the likelihood itself and not just its
gradient, and thus all share this advantage over solving (\ref{ascore}).
Note that one can use randomized algorithms to approximate
$\log\det K$ and thus approximate the loglikelihood directly [\citet{zhangY}].
However, this approximation requires first
taking a power series expansion of $K$ and then applying the randomization
trick to each term in the truncated power series; the examples
presented by
\citet{zhangY} show that the approach does not generally provide a good
approximation to the loglikelihood.
Since the accuracy of the power series approximation to $\log\det K$ depends
on the condition number of $K$, some of the filtering
ideas described by \citet{steinchenanitescufiltering} and used to good
effect in Section \ref{sec4} here
could perhaps be of value for approximating $\log\det K$, but we do not
explore that possibility.
See \citet{aune} for some recent developments on stochastic
approximation of log determinants of positive definite matrices.

Let us consider the four approaches of spectral methods, low-rank
approximations, covariance
tapering and composite likelihood in turn.
Spectral approximations to the likelihood can be fast and accurate
for gridded data [\citet{whittle,guyon,dahlhaus}],
although even for gridded data
they may require some prefiltering to work well [\citet{stein1995}].
In addition, the approximations tend to work less well as the number of
dimensions increase [\citet{dahlhaus}] and thus may be problematic for
space--time
data, especially if the number of spatial dimensions is three.
Spectral approximations have been proposed for ungridded data [\citet{fuentes}],
but they do not work as well as they do for gridded data
from either a statistical or
computational perspective, especially if large
subsets of observations do not form a regular grid.
Furthermore, in contrast to the approach we propose here, there
appears to be no easy way of improving the approximations by doing further
calculations, nor is it clear how to assess the loss of efficiency by using
spectral approximations without a large extra computational burden.

Low-rank approximations, in which the covariance matrix is approximated by
a low-rank matrix plus a diagonal matrix, can greatly reduce the burden
of memory and computation relative to the exact likelihood
[\citet{cressiejohannesson,eidsvik}]. However, for
the kinds of applications we have in mind, in which the diagonal component
of the covariance matrix does not dominate the small-scale variation of the
process, these low-rank approximations tend to work poorly and are not a
viable option [\citet{steinkorea}].

Covariance tapering replaces the covariance matrix of interest by a sparse
covariance matrix with similar local behavior [\citet{furrer}].
There is theoretical support for this approach [\citet{kaufman,wangloh}],
but the tapered covariance matrix must be very sparse to
help a great deal with calculating the log determinant of the covariance
matrix, in which case \citet{steintaper}
finds that composite likelihood approaches
will often be preferable. There is scope for combining covariance tapering
with the approach presented here in that sparse matrices lead to efficient
matrix--vector multiplication, which is also essential for our implementation
of computing (\ref{ascore}) based on iterative methods to do the
matrix solves.
\citet{sang} show that covariance tapering and low-rank approximations can
also sometimes be profitably combined to approximate likelihoods.

We consider methods based on composite likelihoods to be the main competitor
to solving (\ref{ascore}).
The approximate loglikelihoods described by \citet
{vecchia,steinchiwelty,caragea} can all be written in the following form:
for some sequence of pairs of matrices $(A_j,B_j)$, $j=1,\ldots,q$, all with
$n$ columns, at most $n$ rows and full rank,
%
%e1.3 #&#
\begin{equation}\label{composite}
\sum_{j=1}^q \log f_{j,\theta}(A_j
Z\mid B_j Z),
\end{equation}
where $f_{j,\theta}$ is the conditional Gaussian
density of $A_j Z$ given $B_j Z$.
As proposed by \citet{vecchia} and \citet{steinchiwelty}, the rank of $B_j$
will generally be larger than that of $A_j$, in which case the main
computation in
obtaining (\ref{composite}) is
finding Cholesky decompositions of the covariance matrices of
$B_1Z,\ldots,
B_q Z$.
For example, \citet{vecchia} just lets $A_j Z$ be the $j$th component
of $Z$
and $B_j Z$ some subset of $Z_1,\ldots,Z_{j-1}$.
If $m$ is the largest of these subsets, then the
storage requirements for this computation are $O(m^2)$ rather than $O(n^2)$.
Comparable to increasing the number of $U_j$'s in the randomized algorithm
used here, this approach can be updated to obtain a better approximation
of the likelihood by increasing the size of the subset of $Z_1,\ldots,Z_{j-1}$
to condition on when computing the conditional density of $Z_j$.
However, for this approach to be efficient from the perspective of flops,
one needs to store the Cholesky decompositions of the covariance
matrices of
$B_1Z,\ldots,B_q Z$, which would greatly increase the memory requirements
of the algorithm.
For dealing with truly massive data sets, our long-term
plan is to combine the randomized approach studied here with a
composite likelihood by using the randomized
algorithms to compute the gradient of (\ref{composite}), thus making
it possible to consider $A_j$'s and $B_j$'s of larger rank than would
be feasible if one had to do exact calculations.

Section \ref{sec2} provides a bound on the efficiency of the estimating equations
based on the approximate likelihood relative to the Fisher information matrix.
The bound is in terms of the condition number of the true covariance matrix
of the observations and shows that if the covariance matrix is
well conditioned, $N$ does not need to be very large to obtain nearly
optimal estimating equations.
Section \ref{sec3} shows how one can get improved estimating equations
by choosing the $U_j$'s in (\ref{ascore}) based on a design
related to $2^n$ factorial designs.
Section \ref{sec4} describes details of the algorithms, including methods for solving
the approximate score equations and the role of preconditioning.
Section \ref{sec5} provides results of numerical experiments on simulated data.
These results show that the basic method can work well for moderate
values of $N$,
even sometimes when the condition
numbers of the covariance matrices do not stay bounded as the number of
observations increases.
Furthermore, the algorithm with the $U_j$'s chosen as in Section \ref{sec3} can lead
to substantially more accurate approximations for a given $N$.
A large-scale numerical experiment shows that for observations on a partially
occluded grid, the algorithm scales nearly linearly in the sample size.
Section \ref{sec6} applies the methods to OMI (Ozone Monitoring Instrument)
Level 3 (gridded) total column ozone
measurements for April 2012 in the latitude band $40^\circ$--$50^\circ$N.
%
%f1 #&#
\begin{figure}

\includegraphics{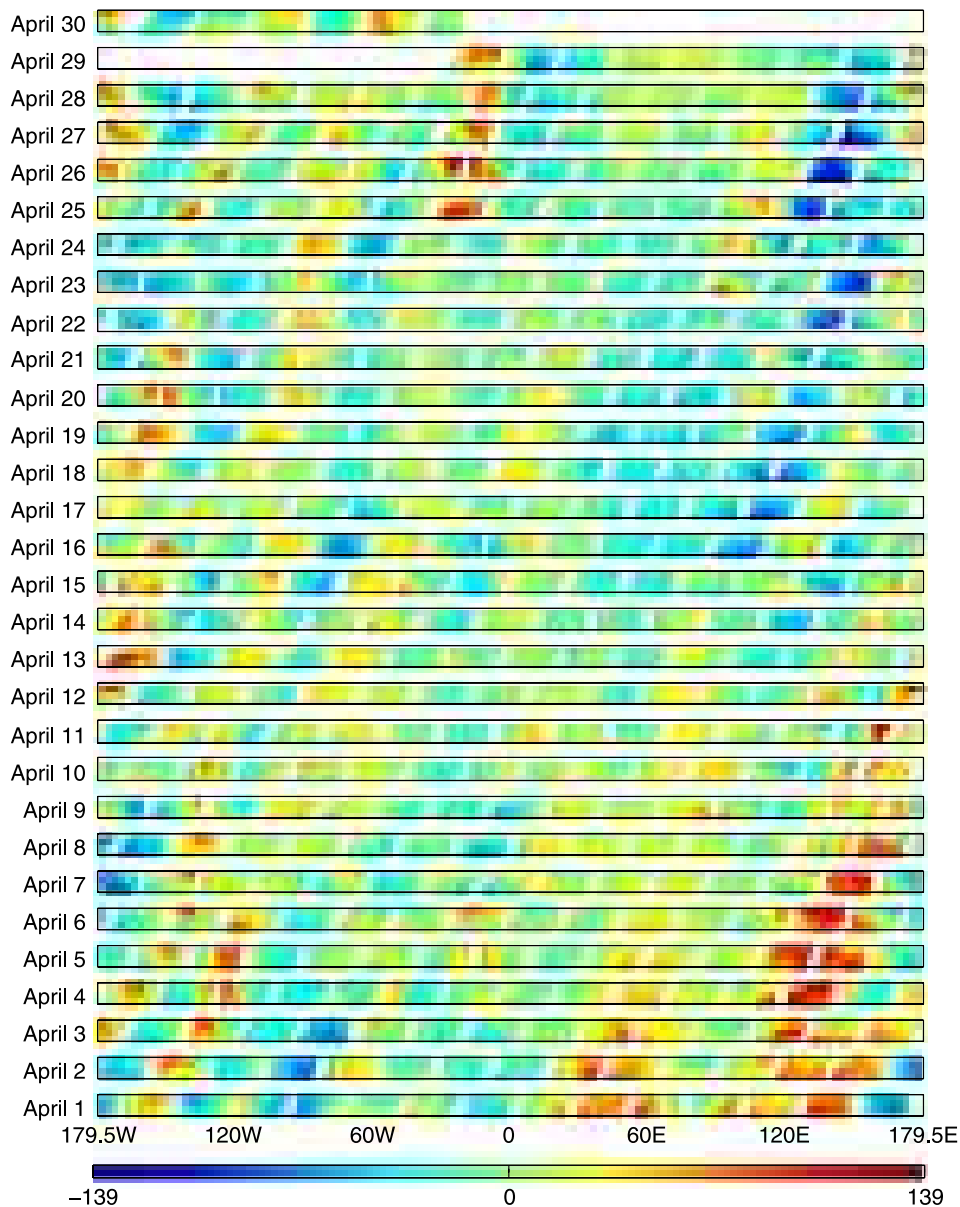}

\caption{Demeaned ozone data (Dobson units)
plotted using a heat color map. Missing data is colored
white.}
\label{figozone}
\end{figure}
The data are given on a $1^\circ\times1^\circ$ grid, so if the data were
complete, there would be a total of $360\times10\times30 = 108\mbox{,}000$
observations.
However, as Figure \ref{figozone} shows, there are missing
observations, mostly due to a lack of overlap in data from different orbits
taken by OMI, but also due to nearly a full day of missing data on April
29--30, so that there are 84,942 observations.
By acting as if all observations are taken at noon local time and assuming
the process is stationary in longitude and time, the covariance matrix
for the
observations can be embedded in a block circulant matrix, greatly reducing
the computational effort needed for multiplying the covariance matrix
by a
vector.
Using (\ref{ascore}) and a factorized
sparse inverse preconditioner [\citet{kolo}], we
are able to compute an accurate approximation to the MLE for a simple model
that captures some of the main features in the OMI data, including the
obvious movement of ozone from day to day visible in
Figure \ref{figozone} that coincides with the prevailing westerly winds
in this latitude band.

%s2 #&#
\section{Variance of stochastic approximation of the score function}\label{sec2}
This section gives a bound relating the covariance matrices
of the approximate and exact score functions.
Let us first introduce some general notation for unbiased estimating
equations.
Suppose $\theta$ has $p$ components and $g(\theta)=(g_1(\theta
),\ldots,
g_p(\theta))'=0$ is a set of
unbiased estimating equations for $\theta$ so that $E_\theta g(\theta)=0$
for all $\theta$.
Write $\dot{g}(\theta)$ for the $p\times p$ matrix whose $ij$th
element is\vspace*{2pt}
$\frac{\partial}{\partial\theta_i}g_j(\theta)$ and $\operatorname{cov}_\theta
\{g(\theta)\}$ for
the covariance matrix of $g(\theta)$.
The Godambe information matrix [\citet{varin}],
\[
\mathcal{E}\bigl\{g(\theta)\bigr\}= E_\theta\bigl\{\dot{g}(\theta)\bigr
\} \bigl[\operatorname{cov}_\theta\bigl\{g(\theta)\bigr\} \bigr]^{-1}
E_\theta\bigl\{\dot{g}(\theta)\bigr\}
\]
is a natural measure of the informativeness of the estimating equations
[\citet{Heyde}, Definition 2.1].
For positive semidefinite matrices $A$ and $B$, write $A\succeq B$ if
$A-B$ is positive semidefinite.
For unbiased estimating equations $g(\theta)=0$ and $h(\theta)=0$,
then we can say $g$ dominates $h$ if $\mathcal{E}\{g(\theta)\}
\succeq\mathcal{E}\{h(\theta)\}$.
Under sufficient regularity conditions on the model and the estimating
equations, the score equations are the optimal estimating equations
[\citet{Bhapkar}].
Specifically,
for the score equations, the Godambe information matrix equals the Fisher
information matrix, ${\cal I}(\theta)$, so this optimality condition
means ${\cal I}(\theta)\succeq\mathcal{E}\{g(\theta)\}$ for all unbiased
estimating equations $g(\theta)=0$.
Writing $M_{ij}$ for the $ij$th element of the matrix $M$,
for the score equations in (\ref{score}),
${\cal I}_{ij}(\theta) = \frac{1}{2}\operatorname{tr}(K^{-1}K_iK^{-1}K_j)$
[\citet{stein-book}, page~179].
For the approximate score equations (\ref{ascore}), it is not difficult
to show that $E_\theta\dot{g}(\theta,N)=-{\cal I}(\theta)$.
Furthermore, writing $W^i$ for $K^{-1}K_i$ and defining the matrix
${\cal J}(\theta)$ by ${\cal J}_{ij}(\theta) =
\operatorname{cov}(U_1'W^iU_1,U_1'W^jU_1)$, we have
%
%e2.1 #&#
\begin{equation}\label{B}
\operatorname{cov}_\theta\bigl\{g(\theta,N)\bigr\} = {\cal I}(\theta) +
\frac{1}{4N}{\cal J}(\theta),
\end{equation}
so that $\mathcal{E}\{g(\theta,N)\}={\cal I}(\theta) \{{\cal
I}(\theta)
+\frac{1}{4N}{\cal J}(\theta) \}^{-1}{\cal I}(\theta)$, which,
as $N\to\infty$, tends to ${\cal I}(\theta)$.

In fact, as also demonstrated empirically
by \citet{anitescu2012mfa}, one may often not need $N$ to be that large
to get estimating
equations that are nearly as efficient as the exact score equations.
Writing $U_{1j}$ for the $j$th component of $U_1$, we have
%
%e2.2 #&#
\begin{eqnarray}\label{Jij}
{\cal J}_{ij}(\theta) & = & \sum_{k,\ell,p,q=1}^n
\operatorname{cov}\bigl(W_{k\ell}^i U_{1k}U_{1\ell},
W_{pq}^j U_{1p}U_{1q}\bigr)
\nonumber
\\
& = & \sum_{k\ne\ell}\bigl\{\operatorname{cov}
\bigl(W_{k\ell}^i U_{1k}U_{1\ell
},W_{k\ell}^j
U_{1k}U_{1\ell}\bigr)+\operatorname{cov}\bigl(W_{k\ell}^i
U_{1k}U_{1\ell},W_{\ell k}^j
U_{1k}U_{1\ell}\bigr)\bigr\}
\nonumber\\[-8pt]\\[-8pt]
& = & \sum_{k\ne\ell}\bigl(W_{k\ell}^i
W_{k\ell}^j + W_{k\ell}^i
W_{\ell k}^j\bigr)
\nonumber
\\
& = & \operatorname{tr}\bigl(W^i W^j\bigr) + \operatorname{tr}
\bigl\{W^i \bigl(W^j\bigr)' \bigr\} - 2\sum
_{k=1}^n W_{kk}^iW_{kk}^j.\nonumber
\end{eqnarray}
As noted by \citet{hutchinson}, the terms with $k=\ell$ drop out
in the second step because $U_{1j}^2=1$ with probability 1. When
$K(\theta)$ is diagonal for all $\theta$, then $N=1$ gives the exact
score equations, although in this case computing
$\operatorname{tr}(K^{-1}K_i)$ directly would be trivial.

Writing $\kappa(\cdot)$ for
the condition number of a matrix,
we can bound\break $\operatorname{cov}_\theta\{g(\theta,N)\}$
in terms of ${\cal I}(\theta)$ and $\kappa(K)$.
The proof of the following result is given in the \hyperref[app]{Appendix}.
%
%th2.1 #&#
\begin{theorem}\label{tmain}
%
%e2.3 #&#
\begin{equation}\label{Bbound}
\operatorname{cov}_\theta\bigl\{g(\theta,N)\bigr\} \preceq{\cal I}(\theta)
\biggl\{ 1 + \frac{(\kappa(K)+1)^2}{4N\kappa(K)} \biggr\}.
\end{equation}
\end{theorem}
It follows from (\ref{Bbound}) that
\[
\mathcal{E}\bigl\{g(\theta,N)\bigr\} \succeq\biggl\{ 1+ \frac{(\kappa
(K)+1)^2}{4N\kappa(K)}
\biggr\}^{-1}{\cal I}(\theta).
\]
In practice, if $\frac{(\kappa(K)+1)^2}{4N\kappa(K)} < 0.01$, so
that the loss
of information in using (\ref{ascore}) rather than (\ref{score}) was
at most
1\%, we would generally be satisfied with using the approximate score
equations and a loss of information of even 10\% or larger might be acceptable
when one has a massive amount of data.
For example, if $\kappa(K)=5$, a~bound of 0.01 is obtained with
$N=180$ and
a bound of 0.1 with $N=18$.

It is possible to obtain unbiased estimating equations similar
to (\ref{ascore}) whose statistical efficiency does not depend on
$\kappa(K)$.
Specifically,
if we write $\operatorname{tr}(K^{-1}K_i)$ as $\operatorname{tr}((G')^{-1}K_iG^{-1})$,
where $G$ is any matrix satisfying $G'G=K$,
we then have that
%
%e2.4 #&#
\begin{equation}\label{symscore}
h_i(\theta,N) = \frac{1}{2}Z'K^{-1}K_iK^{-1}Z
-\frac{1}{2N}\sum_{j=1}^N
U_j'\bigl(G'\bigr)^{-1}K_iG^{-1}U_j
= 0
\end{equation}
for $i=1,\ldots,p$ are also unbiased estimating equations for $\theta$.
In this case,
$\operatorname{cov}_\theta\{h(\theta,N)\}\preceq
( 1+\frac{1}{N} ){\cal I}(\theta)$, whose proof is similar
to that of
Theorem \ref{tmain} but exploits the symmetry of
$(G')^{-1}K_iG^{-1}$. This
bound is less than or equal to the
bound in (\ref{Bbound}) on $\operatorname{cov}_\theta\{g(\theta,N)\}$.
Whether it is preferable to use (\ref{symscore}) rather than (\ref{ascore})
depends on a number of factors, including the sharpness of the bound in
(\ref{Bbound}) and how much more work it takes to compute $G^{-1}U_j$ than
to compute $K^{-1}U_j$. An example of how the action of such a matrix
square root can be approximated efficiently using only $O(n)$ storage is
presented by \citet{chen2011computing}.

%s3 #&#
\section{Dependent designs}\label{sec3}\label{secdependent}
Choosing the $U_j$'s independently is simple and convenient, but
one can reduce the variation in the stochastic approximation by using
a more sophisticated design for the $U_j$'s; this section describes such
a design. Suppose that $n=Nm$ for some nonnegative integer $m$
and that $\beta_1,\ldots,\beta_N$ are fixed
vectors of length $N$ with all entries $\pm1$
for which $\frac{1}{N}\sum_{j=1}^N \beta_j\beta'_j = I$.
For example, if $N=2^q$ for a positive integer $q$, then the $\beta_j$'s
can be chosen to be the design matrix for a saturated model of a $2^q$
factorial design in which the levels of the factors are set at
$\pm1$ [\citet{BHH}, Chapter 5].
In addition, assume that $X_1,\ldots,X_m$ are random
diagonal matrices of size $N$
and $Y_{jk}$, $j=1,\ldots,N; k=1,\ldots,m$ are random variables such
that all the diagonal elements of the $X_j$'s and all the $Y_{jk}$'s
are i.i.d. symmetric Bernoulli random variables.
Then define
%
%e3.1 #&#
\begin{equation}\label{Uj}
U_j = \pmatrix{
Y_{j1}X_1
\cr
\vdots
\cr
Y_{jm}X_m}
\beta_j.
\end{equation}
One can easily show that for any $Nm\times Nm$ matrix $M$,
$E (\frac{1}{N}\sum_{j=1}^N U_j'MU_j ) = \operatorname{tr}(M)$.
Thus, we can use this
definition of the $U_j$'s in (\ref{ascore}), and the resulting estimating
equations are still unbiased.

This design is closely related to a class of designs introduced
by \citet{avron}, who propose selecting the $U_j$'s as follows.
Suppose $H$ is a Hadamard matrix, that is, an $n\times n$ orthogonal
matrix with
elements $\pm1$.
\citet{avron} actually consider $H$ a multiple of a unitary matrix,
but the special case $H$ Hadamard makes their proposal most similar to ours.
Then, using simple random sampling (with replacement), they choose $N$
columns from this matrix and multiply this $n\times N$ matrix by an
$n\times
n$ diagonal matrix with diagonal entries made up of independent symmetric
Bernoulli random variables. The columns of this resulting matrix are
the $U_j$'s.
We are also multiplying a subset of the columns of a Hadamard matrix
by a random diagonal matrix, but we do not select the columns by simple random
sampling from some arbitrary Hadamard matrix.

The extra structure we impose yields beneficial results in terms of the
variance of the randomized trace approximation, as the following calculations
show.
Partitioning $M$ into an $m\times m$ array of
$N\times N$ matrices with
$k\ell$th block $M^b_{k\ell}$, we obtain the following:
%
%e3.2 #&#
\begin{equation}\label{UjMUj}
\frac{1}{N}\sum_{j=1}^N
U_j'MU_j = \frac{1}{N} \sum
_{k,\ell=1}^m\sum_{j=1}^N
Y_{jk}Y_{j\ell} \beta_j'X_k
M^b_{k\ell} X_\ell\beta_j.
\end{equation}
Using $Y_{jk}^2=1$ and $X_k^2 = I$, we have
\begin{eqnarray*}%\label{miracle}
\frac{1}{N}\sum_{j=1}^N
Y_{jk}^2 \beta_j'X_k
M^b_{kk} X_k \beta_j & = &
\frac{1}{N}\operatorname{tr}\Biggl(X_k M^b_{kk}
X_k \sum_{j=1}^N
\beta_j\beta_j' \Biggr)
\\
& = & \operatorname{tr}\bigl(M^b_{kk}X_k^2
\bigr)
\\
& = & \operatorname{tr}\bigl(M^b_{kk}\bigr),
\end{eqnarray*}
which is not random.
Thus, if $M$ is block diagonal (i.e., $M^b_{k\ell}$ is a matrix of
zeroes for
all $k\ne\ell$), (\ref{UjMUj}) yields $\operatorname{tr}(M)$ without error.
This result is an extension of the result that independent $U_j$'s
give $\operatorname{tr}(M)$ exactly for diagonal $M$.
Furthermore, it turns out that, at least in terms of the variance
of $\frac{1}{N}\sum_{j=1}^N U_j'MU_j$, for the elements of $M$ off the
block diagonal, we do exactly the same as we do when
the $U_j$'s are independent.
Write $B(\theta)$ for
$\operatorname{cov}\{g(\theta,N)\}$ with
$g(\theta,N)$ defined as in (\ref{ascore}) with independent $U_j$'s.
Define $g^d(\theta,N)=0$ for
the unbiased estimating equations defined by (\ref{ascore}) with dependent
$U_j$'s defined by (\ref{Uj}) and $B^d(\theta)$ to be the covariance
matrix of
$g^d(\theta,N)$.
Take $T(N,n)$
to be the set of pairs of
positive integers $(k,\ell)$ with
$1\le\ell<k \le n$ for which $\lfloor k/N\rfloor= \lfloor\ell
/N\rfloor$.
We have the following result, whose proof is given in the \hyperref[app]{Appendix}.

%th3.1 #&#
\begin{theorem}\label{tdependent}
For any vector $v=(v_1,\ldots,v_p)'$,
%
%e3.3 #&#
\begin{equation}\label{improve}
v'B(\theta)v - v'B^d(\theta)v =
\frac{2}{N} \sum_{(k,\ell)\in T(N,n)} \Biggl\{ \sum
_{i=1}^p v_i \bigl( W_{k\ell}^i+W_{\ell k}^i
\bigr) \Biggr\}^2.
\end{equation}
\end{theorem}
Thus, $B(\theta) \succeq B^d(\theta)$.
Since $E_\theta\dot{g}
(\theta,N) = E_\theta\dot{g}^d(\theta,N)=-\mathcal{I}(\theta)$, it
follows that
$\mathcal{E}\{g^d(\theta,N)\}\succeq\mathcal{E}\{g(\theta,N)\}$.

How much of an improvement will result from using dependent $U_j$'s depends
on the size of the $W_{k\ell}^i$'s within each block.
For spatial data, one would typically group spatially contiguous observations
within blocks.
How to block for space--time data is less clear.
The results here focus on the variance of the randomized trace approximation.
\citet{avron} obtain bounds on the probability that the
approximation error is less than some quantity and note that these results
sometimes give rankings for various randomized trace
approximations different from those obtained by comparing variances.

%s4 #&#
\section{Computational aspects}\label{sec4}
Finding $\theta$ that solves
the estimating equations (\ref{ascore}) requires a
nonlinear equation solver in addition to
computing linear solves in $K$. The nonlinear
solver starts at an initial guess $\theta^0$ and iteratively updates
it to
approach a (hopefully unique)
zero of (\ref{ascore}). In each iteration, at $\theta^i$, the
nonlinear solver typically requires an evaluation of $g(\theta^i,N)$
in order
to find the next iterate~$\theta^{i+1}$. In turn, the evaluation of $g$
requires employing a linear solver to compute the set of vectors
$K^{-1}Z$ and
$K^{-1}U_j$, $j=1,\ldots,N$.

The Fisher information matrix $\mathcal{I}(\theta)$ and the matrix
$\mathcal{J}(\theta)$ contain terms involving matrix traces and diagonals.
Write $\diag(\cdot)$ for a column vector containing the diagonal
elements of a matrix and $\circ$ for the Hadamard (elementwise) product
of matrices.
For any real matrix $A$,
\[
\tr(A)=E_U\bigl(U'AU\bigr) \quad\mbox{and}\quad
\diag(A)=E_U(U\circ AU),
\]
where the expectation $E_U$ is taken over $U$, a random vector with
i.i.d.
symmetric Bernoulli components.
One can unbiasedly
estimate $\mathcal{I}(\theta)$ and $\mathcal{J}(\theta)$ by
%
%e4.1 #&#
\begin{equation}\label{Iij}
\hat{\mathcal{I}}_{ij}(\theta)=\frac{1}{2N_2}\sum
_{k=1}^{N_2}U_k'W^iW^jU_k
\end{equation}
and
%
%e4.2 #&#
\begin{eqnarray}
\label{Jij2} \hat{\mathcal{J}}_{ij}(\theta) & = & \frac{1}{N_2}\sum
_{k=1}^{N_2}U_k'W^iW^jU_k
+\frac{1}{N_2}\sum_{k=1}^{N_2}U_k'W^i
\bigl(W^j\bigr)'U_k
\nonumber\\[-8pt]\\[-8pt]
&&{} -2\sum_{\ell=1}^n \Biggl[
\frac{1}{N_2}\sum_{k=1}^{N_2}
\bigl(U_k\circ W^iU_k\bigr)
\Biggr]_{\ell} \Biggl[\frac{1}{N_2}\sum_{k=1}^{N_2}
\bigl(U_k\circ W^jU_k\bigr)
\Biggr]_{\ell}.
\nonumber
\end{eqnarray}
Note that here the set of vectors $U_k$ need not be the same as that
in (\ref{ascore}) and that $N_2$ may not be the same as $N$, the number
of $U_j$'s used to compute the estimate of~$\theta$.
Evaluating $\hat{\mathcal{I}}(\theta)$ and $\hat{\mathcal
{J}}(\theta)$
requires linear solves since $W^iU_k=K^{-1}(K_iU_k)$ and
$(W^i)'U_k=K_i(K^{-1}U_k)$. Note that one can also unbiasedly estimate
$\mathcal{J}_{ij}(\theta)$ as the sample covariance of $U'_kW^iU_k$ and
$U'_kW^jW_k$ for $k=1,\ldots,N$, but (\ref{Jij2}) directly exploits
properties of symmetric Bernoulli variables (e.g., $U^2_{1j}=1$).
Further study
would be needed to see when each approach is preferred.

%%%%%%%%%%%%%%%%%%%%%%%%%%%%%%%%%%%%%%%%%%%%%%%%%%%%%%%%%%
%s4.1 #&#
\subsection{Linear solver}\label{sec4.1}
We consider an iterative solver for solving a set of linear
equations $Ax=b$ for a symmetric positive definite matrix
$A\in\R^{n\times n}$, given a right-hand vector $b$.
Since the matrix $A$ (in our case the covariance matrix) is symmetric positive
definite, the conjugate gradient algorithm is naturally used. Let $x^i$
be the
current approximate solution, and let $r^i=b-Ax^i$ be the residual. The
algorithm
finds a search direction $q^i$ and a step size $\alpha^i$ to update the
approximate solution, that is, $x^{i+1}=x^i+\alpha^iq^i$, such that the
search directions $q^i,\ldots,q^0$ are mutually $A$-conjugate [i.e.,
$(q^i)'Aq^j = 0$ for $i\ne j$] and the new
residual $r^{i+1}$ is orthogonal to all the previous ones, $r^i,\ldots,r^0$.
One can show that the search direction is a linear combination of the current
residual and the past search direction, yielding the following recurrence
formulas:
\begin{eqnarray*}
x^{i+1}&=&x^i+\alpha^iq^i,
\\
r^{i+1}&=&r^i-\alpha^iAq^i,
\\
q^{i+1}&=&r^{i+1}+\beta^iq^i,
\end{eqnarray*}
where $\alpha^i= \langle r^i,r^i \rangle/ \langle
Aq^i,q^i \rangle$ and
$\beta^i= \langle r^{i+1},r^{i+1} \rangle/ \langle
r^i,r^i \rangle$, and
$ \langle\cdot,\cdot\rangle$ denotes the vector inner
product. Letting
$x^*$ be the
exact solution, that is, $Ax^*=b$, then $x^i$ enjoys a linear
convergence to
$x^*$:
%
%e4.3 #&#
\begin{equation}
\label{eqncgconverge} \bigl\|x^i-x^*\bigr\|_A\le2 \biggl(
\frac{\sqrt{\kappa(A)}-1}{\sqrt{\kappa
(A)}+1} \biggr)^{i}\bigl\|x^0-x^*\bigr\|_A,
\end{equation}
where $\|\cdot\|_A= \langle A\cdot,\cdot\rangle^{{1/2}}$ is the
$A$-norm of a vector.

Asymptotically, the time cost of one iteration is upper
bounded by that of multiplying $A$ by $q^i$, which typically
dominates other vector operations when $A$ is not sparse. Properties of the
covariance matrix can be exploited to efficiently compute
the matrix--vector products. For example, when the
observations are on a lattice (regular grid), one can use
the fast Fourier transform (FFT), which takes time $O(n\log
n)$ [\citet{toeplitzbook}]. Even when the grid is partial
(with occluded observations), this idea can still be
applied. On the other hand, for nongridded observations,
exact multiplication generally requires $O(n^2)$ operations.
However,
one can use a combination of direct summations for close-by
points and multipole expansions of the covariance kernel for
faraway points to compute the matrix--vector products in
$O(n\log n)$, even $O(n)$, time [\citet{treecode,fmm}]. In
the case of Mat\'ern-type Gaussian processes and in the
context of solving the stochastic
approximation (\ref{ascore}), such fast
multipole approximations were presented
by \citet{anitescu2012mfa}. Note that the total computational cost of
the solver is the cost of each iteration times the number of
iterations, the latter being usually much less than $n$.

The number of iterations to achieve a desired accuracy depends on how
fast $x^i$ approaches $x^*$, which, from (\ref{eqncgconverge}), is in
turn affected by the condition number $\kappa$ of~$A$. Two techniques
can be used to improve convergence. One is to perform preconditioning
in order to reduce $\kappa$; this technique will be discussed in the
next section. The other is to adopt a block version of the conjugate
gradient algorithm. This technique is useful for solving the linear
system for the same matrix with multiple right-hand sides.
Specifically, denote by $AX=B$ the linear system one wants to solve,
where $B$ is a matrix with $s$ columns, and the same for the
unknown~$X$. Conventionally, matrices such as $B$ are called
\textit{block vectors}, honoring the fact that the columns of $B$ are
handled simultaneously. The block conjugate gradient algorithm is
similar to the single-vector version except that the iterates $x^i$,
$r^i$ and $q^i$ now become block iterates $X^i$, $R^i$ and $Q^i$ and
the coefficients $\alpha^i$ and $\beta^i$ become $s\times s$ matrices.
The detailed algorithm is not shown here; interested readers are
referred to \citet{olearyblockcg}. If $X^*$ is the exact solution,
then $X^i$ approaches $X^*$ at least as fast as linearly:
%
%e4.4 #&#
\begin{equation}
\label{eqnbcgconverge} \bigl\|\bigl(X^i\bigr)_j-\bigl(X^*
\bigr)_j\bigr\|_A\le C_j \biggl(\frac{\sqrt{\kappa_s(A)}-1}{\sqrt{\kappa_s(A)}+1}
\biggr)^{i},\qquad j=1,\ldots,s,
\end{equation}
where $(X^i)_j$ and $(X^*)_j$ are the $j$th column of $X^i$ and $X^*$,
respectively; $C_j$ is some constant dependent on $j$ but not $i$; and
$\kappa_s(A)$ is the ratio between $\lambda_n(A)$ and $\lambda_s(A)$
with the
eigenvalues $\lambda_k$ sorted increasingly. Comparing (\ref
{eqncgconverge}) with (\ref{eqnbcgconverge}), we see that the
modified condition number $\kappa_s$ is less than $\kappa$, which
means that the block version of the conjugate gradient algorithm has a
faster convergence than the standard version does. In practice, since
there are many right-hand sides (i.e., the vectors $Z$, $U_j$'s and
$K_iU_k$'s), we always use the block version.

%%%%%%%%%%%%%%%%%%%%%%%%%%%%%%%%%%%%%%%%%%%%%%%%%%%%%%%%%%
%s4.2 #&#
\subsection{Preconditioning/filtering}\label{sec4.2}
Preconditioning is a technique for reducing the condition number of the
matrix. Here, the benefit of preconditioning is twofold: it encourages
the rapid convergence of an iterative linear solver and, if the
effective condition number is small, it strongly bounds the uncertainty
in using the estimating equations (\ref{ascore}) instead of the exact
score equations (\ref{score}) for estimating parameters (see
Theorem \ref{tmain}). In numerical linear algebra, preconditioning
refers to applying a matrix $M$, which approximates the inverse of $A$
in some sense, to both sides of the linear system of equations. In the
simple case of left preconditioning, this amounts to solving $MAx=Mb$
for $MA$ better conditioned than $A$. With certain algebraic
manipulations, the matrix $M$ enters into the conjugate gradient
algorithm in the form of multiplication with vectors. For the detailed
algorithm, see \citet{saadbookiterativemethod}. This technique does
not explicitly compute the matrix $MA$, but it requires that the
matrix--vector multiplications with $M$ can be efficiently carried
out.\looseness=-1

For covariance matrices, certain filtering operations are known to
reduce the
condition number, and some can even achieve an optimal preconditioning
in the
sense that the condition number is bounded by a constant independent of the
size of the matrix [\citet{steinchenanitescufiltering}]. Note that these
filtering operations may or may not preserve the rank/size of the matrix.
When the rank is reduced, then some loss of statistical
information results when filtering, although similar filtering is also
likely needed to apply spectral methods for strongly correlated spatial
data on a grid [\citet{stein1995}].
Therefore, we consider applying the same filter to all the vectors and
matrices in the estimating equations, in which case
(\ref{ascore}) becomes the
stochastic approximation to the score equations of the \textit{filtered}
process. Evaluating the filtered version of $g(\theta,N)$ becomes easier
because the linear solves with the filtered covariance matrix converge
faster.

%%%%%%%%%%%%%%%%%%%%%%%%%%%%%%%%%%%%%%%%%%%%%%%%%%%%%%%%%%
%s4.3 #&#
\subsection{Nonlinear solver}\label{sec4.3}\label{secnonlinearsolver}
The choice of the nonlinear solver is problem dependent. The purpose of
solving the score equations (\ref{score}) or the estimating
equations (\ref{ascore}) is to maximize the loglikelihood function
$\mathcal{L}(\theta)$. Therefore, investigation into the shape of the
loglikelihood surface helps identify an appropriate solver.

In Section \ref{sec5}, we consider the power law generalized covariance model
($\alpha>0$):
%
%e4.5 #&#
\begin{equation}
\label{GC1} G(x;\theta)= \cases{\Gamma(-\alpha/2) r^{\alpha}, &\quad if $
\alpha/2\notin\N$,
\cr
(-1)^{1+\alpha/2}r^{\alpha}\log r, &\quad if $\alpha/2\in
\N$,}
\end{equation}
where $x=[x_1,\ldots,x_d]\in\R^d$ denotes coordinates, $\theta$ is the
set of parameters containing $\alpha>0$,
$\ell=[\ell_1,\ldots,\ell_d]\in\R^d$, and $r$ is the elliptical radius
%
%e4.6 #&#
\begin{equation}\label{GC2}
r=\sqrt{\frac{x_1^2}{\ell_1^2}+\cdots+\frac{x_d^2}{\ell_d^2}}.
\end{equation}
Allowing a different scaling in different directions may be appropriate when,
for example, variations in a vertical direction may be different from
those in a
horizontal direction.
The function $G$ is conditionally positive definite; therefore, only the
covariances of authorized linear combinations of the process are
defined [\citet{geostatisticsbook}, Section 4.3].
In fact, $G$ is $p$-conditionally positive definite
if and only if $2p+2>\alpha$ [see \citet{geostatisticsbook},
Section 4.5], so that
applying the discrete Laplace filter (which gives second-order differences)
$\tau$ times to the observations
yields a set of authorized linear combinations when $\tau\ge\frac
{1}{2}\alpha$.
\citet{steinchenanitescufiltering} show that
if $\alpha=4\tau-d$, then the
covariance matrix has a bounded condition number independent of the matrix
size.
Consider the grid $\{\delta\mathbf{j}\}$
for some fixed spacing $\delta$ and $\mathbf{j}$ a vector whose components
take integer values between $0$ and $m$.
Applying the filter $\tau$ times, we obtain the covariance matrix
\[
K_{\mathbf{i}\mathbf{j}}=\cov\bigl\{\Delta^{\tau}Z(\delta\mathbf{i}),\Delta
^{\tau}Z(\delta\mathbf{j})\bigr\},
\]
where $\Delta$ denotes the discrete Laplace operator
\[
\Delta Z(\delta\mathbf{j})=\sum_{p=1}^d\bigl
\{Z(\delta\mathbf{j}-\delta\mathbf{e}_p)-2Z(\delta\mathbf{j})+Z(\delta\mathbf{j}+
\delta\mathbf{e}_p)\bigr\}
\]
with $\mathbf{e}_p$ meaning the unit vector along the $p$th coordinate.
If $\tau=\operatorname{round}((\alpha+d)/4)$,
the resulting $K$ is both positive definite and reasonably well conditioned.

\begin{figure}

\includegraphics{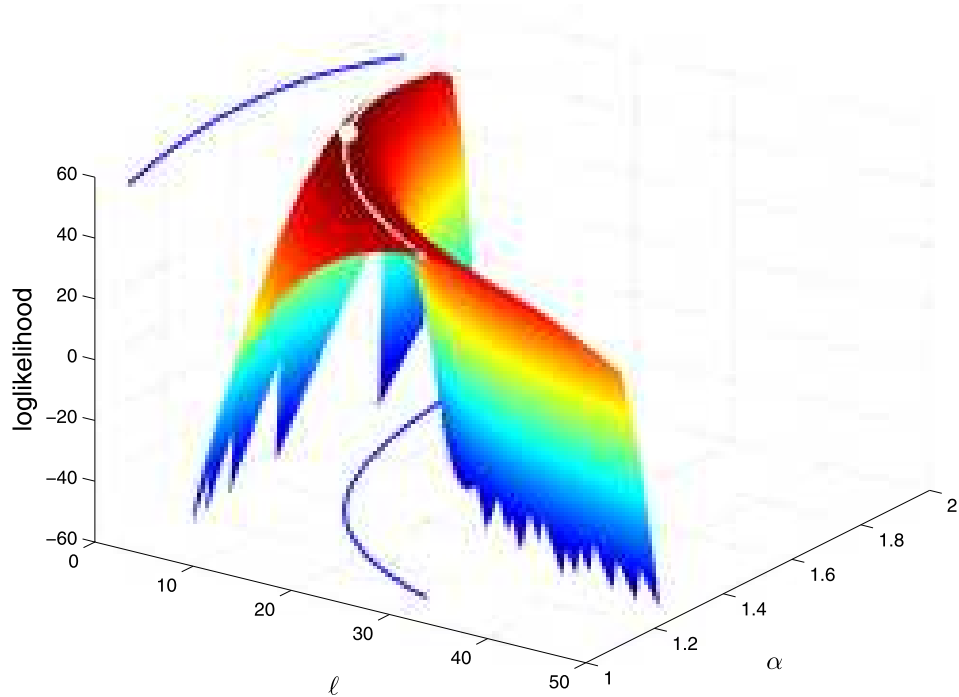}

\caption{A sample loglikelihood surface for the power law generalized
covariance kernel, with profile curve and peak plotted.}
\label{figloglik}
\end{figure}

Figure \ref{figloglik} shows a sample loglikelihood surface for $d=1$ based
on an observation vector $Z$ simulated from a 1D partial regular grid spanning
the range $[0,100]$, using parameters $\alpha=1.5$ and $\ell=10$. (A similar
2D grid is shown later in Figure~\ref{figgrid}.) The peak of the
surface is
denoted by the solid white dot, which is not far away from the truth
$\theta=(1.5,10)$. The white dashed curve (profile of the surface) indicates
the maximum loglikelihoods $\mathcal{L}$ given $\alpha$. The curve is also
projected on the $\alpha-\mathcal{L}$ plane and the $\alpha-\ell$
plane. One
sees that the loglikelihood value has small variation (ranges from $48$ to
$58$) along this curve compared with the rest of the surface, whereas, for
example, varying just the
parameter $\ell$ changes the loglikelihood substantially.

%f2 #&#

A Newton-type nonlinear solver starts at some initial point $\theta^0$ and
tries to approach the optimal point (one that solves the score
equations).\setcounter{footnote}{3}\footnote{To facilitate understanding, we explain here the process
for solving the score equations (\ref{score}). Conceptually it is similar
to that for solving the estimating equations (\ref{ascore}).} Let the current
point be $\theta^i$.\vadjust{\goodbreak} The solver finds a direction $q^i$
and a step size $\alpha^i$ in
some way to move the point to $\theta^{i+1}=\theta^i+\alpha^iq^i$,
so that the value of $\mathcal{L}$ is increased. Typically, the search
direction $q^i$ is the inverse of the Jacobian multiplied by $\theta
^i$, that is,
$q^i=\dot{g}(\theta^i,N)^{-1}\theta^i$. This way, $\theta^{i+1}$ is
closer to a solution of the score equations.
Figure \ref{figloglik} shows a loglikelihood surface when $d=1$.
The solver starts somewhere on the
surface and quickly climbs to a point along the profile curve. However, this
point might be far away from the peak. It turns out that along this
curve a
Newton-type solver is usually unable to find a direction with an
appropriate step size to numerically increase $\mathcal{L}$, in part because
of the narrow ridge indicated in the figure.
The variation of $\mathcal{L}$ along
the normal direction of the curve is much larger than that along the tangent
direction. Thus, the iterate $\theta^i$ is trapped and cannot advance
to the
peak. In such a case, even though the estimated maximized likelihood
could be fairly close to the true maximum, the estimated parameters
could be quite distant from the MLE of $(\alpha,\ell)$.

To successfully solve the estimating equations, we consider each
component of $\ell$ an implicit function of $\alpha$. Denote by
%
%e4.7 #&#
\begin{equation}
\label{ascore2} g_i(\alpha,\ell_1,\ldots,
\ell_d)=0,\qquad i=1,\ldots,d+1,\vadjust{\goodbreak}
\end{equation}
the estimating equations, ignoring the fixed variable $N$. The implicit
function theorem indicates that a set of functions
$\ell_1(\alpha),\ldots,\ell_d(\alpha)$ exists around an isolated zero
of (\ref{ascore2}) in a neighborhood where (\ref{ascore2}) is
continuously differentiable, such that
\[
g_i\bigl(\alpha,\ell_1(\alpha),\ldots,
\ell_d(\alpha)\bigr)=0\qquad \mbox{for } i=2,\ldots,d+1.
\]
Therefore, we need only to solve the equation
%
%e4.8 #&#
\begin{equation}
\label{ascore3} g_1\bigl(\alpha,\ell_1(\alpha),\ldots,
\ell_d(\alpha)\bigr)=0
\end{equation}
with a single variable $\alpha$. Numerically, a much more robust method
than a Newton-type method exists for finding a root of a one-variable
function. We use the standard method of Forsythe, Malcolm and Moler
[(\citeyear{fzero}), see the Fortran code \mbox{\texttt{Zeroin}}] for
solving (\ref{ascore3}). This method in turn requires the evaluation of
the left-hand side of (\ref{ascore3}). Then, the $\ell_i$'s are
evaluated by solving $g_2,\ldots,g_{d+1}=0$ fixing $\alpha$, whereby a
Newton-type algorithm is empirically proven to be an efficient method.

%s5 #&#
\section{Experiments}\label{sec5}\label{secexp}
\begin{figure}

\includegraphics{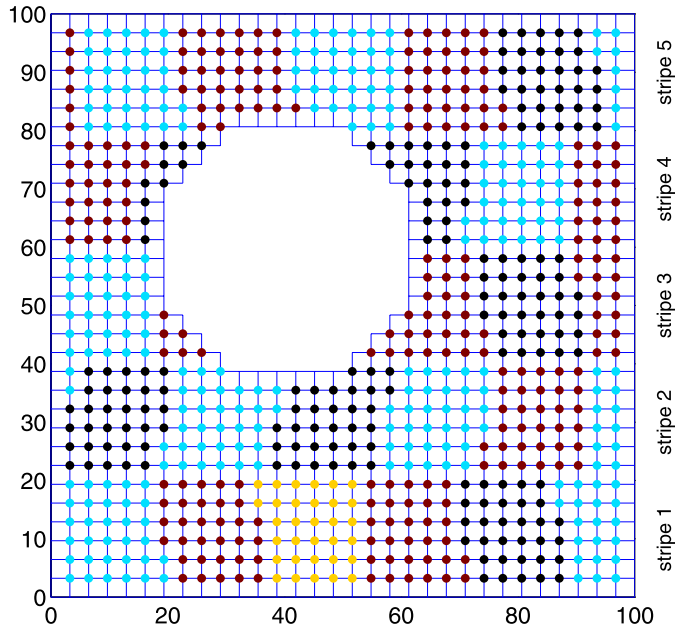}

\caption{A $32\times32$ grid with a region of missing observations in
a disc shape. Internal grid points are grouped to work with the
dependent design in Section \protect\ref{secdependent}.}
\label{figgrid}
\end{figure}

In this section we show a few experimental results based on a
partially occluded regular grid.
The rationale for using
such a partial grid is to illustrate a setting where spectral
techniques do not work so well but
efficient matrix--vector multiplications are available.
A partially occluded grid can occur, for example, when observations
of some surface characteristics are
taken by a satellite-based instrument and it is not possible to obtain
observations over regions with sufficiently dense cloud cover.
The ozone example in Section \ref{sec6} provides another example in which data
on a
partial grid occurs.
This section considers a grid with physical range
$[0,100]\times[0,100]$ and a hole in a disc shape of radius $10$
centered at
$(40,60)$. An illustration of the grid, with size $32\times32$, is
shown in
Figure~\ref{figgrid}. The matrix--vector multiplication is performed
by first
doing the multiplication using the full grid via circulant embedding
and FFT,
followed by removing the entries corresponding to the hole of the grid. Recall
that the covariance model is defined in Section \ref{secnonlinearsolver},
along with the explanation of the filtering step.

%f3 #&#

When working with dependent samples, it is advantageous to group nearby grid
points such that the resulting blocks have a plump shape and that there
are as many blocks with size exactly $N$ as possible. For an
occluded grid, this is a nontrivial task. Here we use a simple
heuristic to
effectively group the points. We divide\vspace*{1pt} the grid into horizontal
stripes of
width $\lfloor\sqrt{N}\rfloor$ (in case $\lfloor\sqrt{N}\rfloor$
does not
divide the grid size along the vertical direction, some stripes have a width
$\lfloor\sqrt{N}\rfloor+1$). The stripes are ordered from bottom to
top, and
the grid points inside the odd-numbered
stripes are ordered lexicographically in their
coordinates, that is, $(x,y)$. In order to obtain as many contiguous
blocks as
possible, the grid points inside the even-numbered stripes are ordered
lexicographically according to $(-x,y)$. This ordering gives a zigzag
flow of the
points starting from the bottom-left corner of the grid. Every $N$
points are
grouped in a block. The coloring of the grid points in Figure \ref{figgrid}
shows an example of the grouping. Note that because of filtering, observations
on either an external or internal boundary are not part of any block.

%%%%%%%%%%%%%%%%%%%%%%%%%%%%%%%%%%%%%%%%%%%%%%%%%%%%%%%%%%
%s5.1 #&#
\subsection{Choice of $N$}\label{sec5.1}
One of the most important factors that affect the efficacy of
approximating the score equations is the value $N$. Theorem \ref{tmain}
indicates that $N$ should increase at least like $\kappa(K)$ in order
to guarantee the additional uncertainty introduced by approximating the
score equations be comparable with that caused by the randomness of the
sample $Z$. In the ideal case, when the condition number of the matrix
(possibly with filtering) is bounded independent of the matrix
size~$n$, then even taking $N=1$ is sufficient to obtain estimates with
the same rate of convergence as the exact score equations. When
$\kappa$ grows with $n$, however, a better guideline for selecting $N$
is to consider the growth of $\mathcal{I}^{-1}\mathcal{J}$.

\begin{figure}

\includegraphics{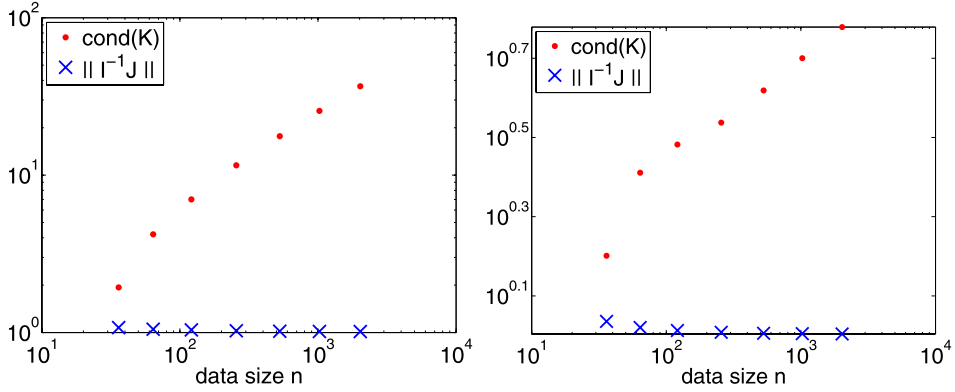}

\caption{Growth of $\kappa$ compared with that of $\|\mathcal
{I}^{-1}\mathcal{J}\|$, for power law kernel in 2D. Left: $\alpha=1$;
right: $\alpha=1.5$.}
\label{figpowerbounds}
\end{figure}

Figure \ref{figpowerbounds} plots the condition number of $K$ and
the spectral norm of
$\mathcal{I}^{-1}\mathcal{J}$ for varying sizes of the matrix and
preconditioning using the Laplacian filter. Although
performing a Laplacian filtering will yield provably bounded
condition numbers only for
the case $\alpha=2$, one sees that the filtering is also effective for the
cases $\alpha=1$ and $1.5$. Moreover, the norm of $\mathcal
{I}^{-1}\mathcal{J}$
is significantly smaller than $\kappa$ when $n$ is large and, in fact,
it does
not seem to grow with $n$. This result indicates the bound in Theorem 1 is
sometimes far too conservative and that using a fixed $N$ can
be effective even when $\kappa$ grows with $n$.

%f4 #&#

\begin{figure}[b]

\includegraphics{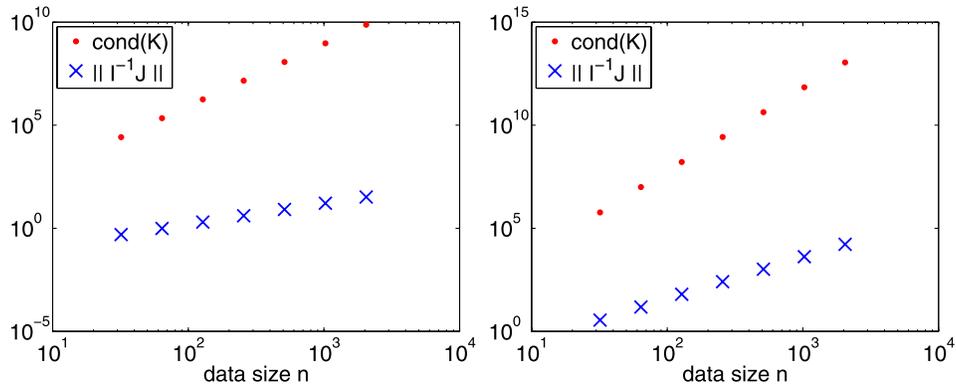}

\caption{Growth of $\kappa$ compared with that of $\|\mathcal
{I}^{-1}\mathcal{J}\|$, for Mat\'{e}rn kernel in 1D, without
filtering. Left: $\nu=1$; right: $\nu=1.5$.}
\label{figmaternbounds}
\end{figure}

Of course, the norm of $\mathcal{I}^{-1}\mathcal{J}$ is not
always bounded. In Figure \ref{figmaternbounds} we show
two examples using the Mat\'{e}rn covariance kernel with
smoothness parameter $\nu=1$ and 1.5 (essentially $\alpha=2$ and 3).
Without filtering, both
$\kappa(K)$ and $\|\mathcal{I}^{-1}\mathcal{J}\|$ grow with
$n$, although the plots show that the growth of the latter
is significantly slower than that of the former.

%f5 #&#

If the occluded observations are more scattered, then the fast matrix--vector
multiplication based on circulant embedding still works fine.
However,
if the occluded pixels are randomly located and the fraction of occluded
pixels is substantial, then using a filtered data set
only including Laplacians centered at those
observations whose four nearest neighbors are also available might lead
to an unacceptable loss of information.
In this case, one might instead use a preconditioner based on a sparse
approximation to the inverse Cholesky decomposition as described in
Section \ref{sec6}.

%%%%%%%%%%%%%%%%%%%%%%%%%%%%%%%%%%%%%%%%%%%%%%%%%%%%%%%%%%
%s5.2 #&#
\subsection{\texorpdfstring{A $32\times32$ grid example}{A 32x32 grid example}}\label{sec5.2}
Here, we show the details of solving the estimating equations (\ref{ascore})
using a $32\times32$ grid as an example. Setting the truth $\alpha
=1.5$ and
$\ell=(7,10)$ [i.e., $\theta=(1.5, 7, 10)$], consider exact and
approximate maximum likelihood estimation based on the data obtained by
applying the Laplacian filter once to the observations.
Writing $\mathcal{G}$ for $\mathcal{E}\{g(\theta,N)\}$,
one way to evaluate the approximate
MLEs is to compute the ratios
of the square roots of the diagonal elements of $\mathcal{G}^{-1}$
to the square roots
of the diagonal elements of $\mathcal{I}^{-1}$.
We know these ratios must be at least 1, and that the closer they are
to 1,
the more nearly optimal the resulting estimating equations based on
the approximate score function are.
For $N=64$ and independent sampling, we get 1.0156, 1.0125 and 1.0135 for
the three ratios, all of which are very close to 1.
Since one generally cannot calculate $\mathcal{G}^{-1}$ exactly, it
is also worthwhile to compare a stochastic approximation of the diagonal
values of $\mathcal{G}^{-1}$ to their exact values.
When this approximation was done once for $N=64$ and by using $N_2=100$
in (\ref{Iij}) and (\ref{Jij2}), the three ratios obtained were 0.9821,
0.9817 and 0.9833, which are all close to 1.

%f6 #&#
\begin{figure}

\includegraphics{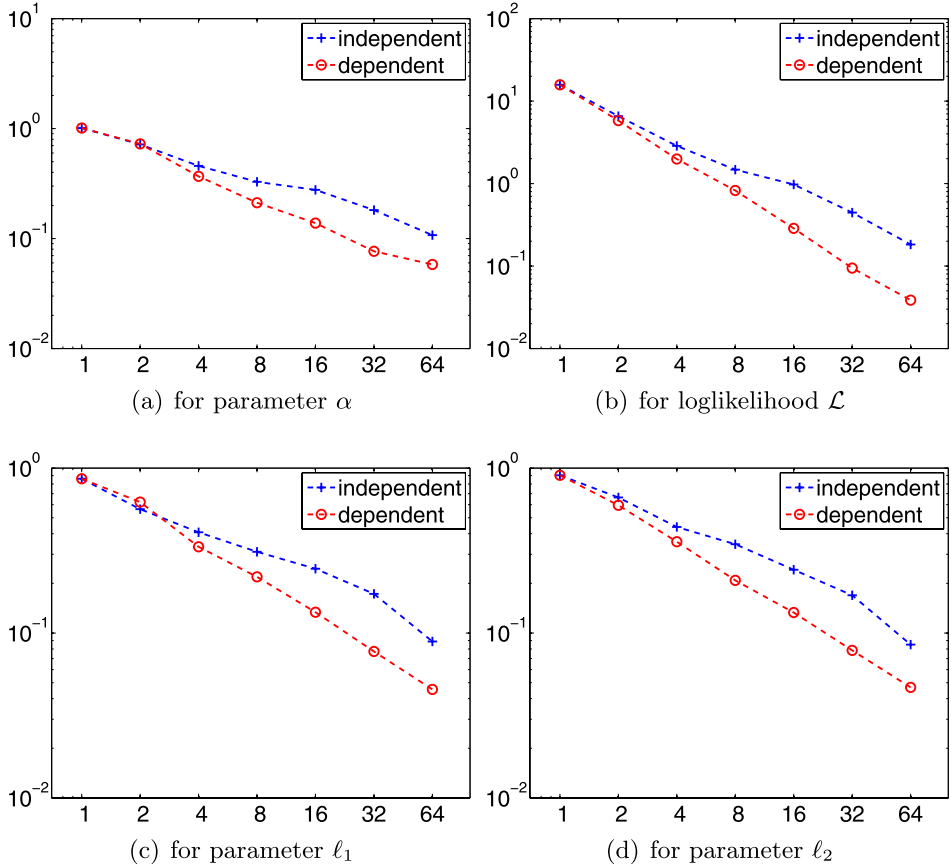}

\caption{Effects of $N$ (1, 2, 4, 8, 16, 32, 64). In each plot, the
curve with
the plus sign corresponds to the independent design, whereas that with the
circle sign corresponds to the dependent design. The horizontal axis
represents $N$. In plots \textup{(a)}, \textup{(c)} and \textup{(d)}, the vertical axis represents the
mean squared differences between the approximate and exact MLEs divided
by the
mean squared errors for the exact MLEs, for the components $\alpha$,
$\ell_1$
and $\ell_2$, respectively. In plot \textup{(b)}, the vertical axis represents
the mean
squared difference between the loglikelihood values at the exact and
approximate MLEs.}
\label{figN}
\end{figure}

Figure \ref{figN} shows the performance of the resulting estimates
(to be compared with the exact MLEs obtained by solving the standard
score equations).
For $N=1$, $2$, $4$, $8$, $16$, $32$
and $64$, we simulated 100 realizations of the process on the $32\times32$
occluded
grid, applied the discrete Laplacian once, and then computed exact
MLEs and approximations using both independent and dependent
(as described in the beginning of Section \ref{secexp}) sampling.
When $N=1$, the independent and dependent
sampling schemes are identical, so only results for independent
sampling are
given. Figure \ref{figN} plots, for each component of $\theta$, the
mean squared
differences between the approximate and exact MLEs
divided by the mean squared errors for the exact MLEs. As expected, these
ratios decrease with $N$, particularly for dependent sampling. Indeed,
dependent sampling is much more efficient than independent sampling for
larger $N$; for example, the results in Figure \ref{figN} show that
dependent sampling with $N=32$ yields better estimates for all three
parameters than does independent sampling with $N=64$.

%%%%%%%%%%%%%%%%%%%%%%%%%%%%%%%%%%%%%%%%%%%%%%%%%%%%%%%%%%
%s5.3 #&#
\subsection{Large-scale experiments}\label{sec5.3}
\begin{figure}[b]

\includegraphics{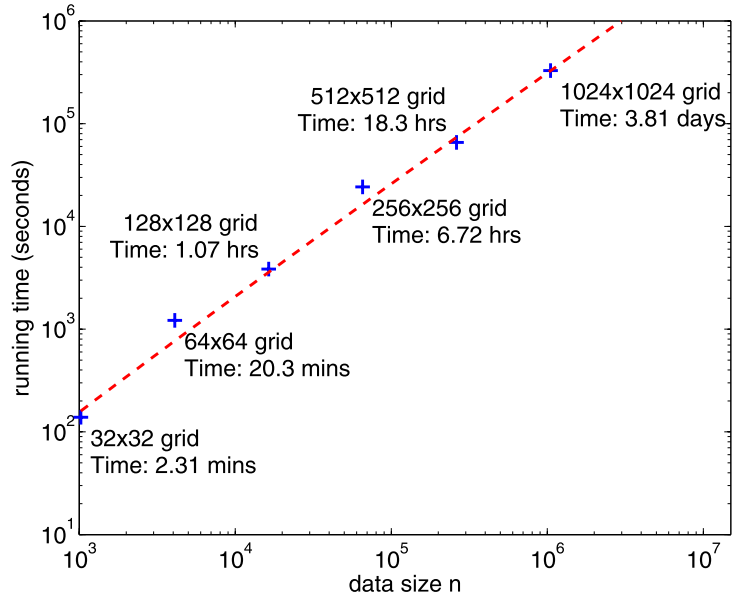}

\caption{Running time for increasingly dense grids. The dashed curve
fits the recorded times with a function of the form of $n\log n$ times a
constant.}
\label{figlargescale}
\end{figure}

We experimented with larger grids (in the same physical range). We show the
results in Table \ref{tablargescale} and Figure \ref
{figlargescale} for
$N=64$. When the matrix becomes large, we are unable to compute
$\mathcal{I}$ and $\mathcal{G}$ exactly. Based on the preceding
experiment, it seems
reasonable to use $N_2=100$ in approximating $\mathcal{I}$ and
$\mathcal{G}$.
Therefore, the elements of $\mathcal{I}$ and $\mathcal{G}$
in Table \ref{tablargescale} were computed only
approximately.

%t1 #&#
\begin{table}
\caption{Estimates and estimated standard errors for increasingly
dense grids.
The last three rows show the ratio of standard errors of the
approximate to the
exact MLEs}
\label{tablargescale}
\begin{tabular*}{\tablewidth}{@{\extracolsep{\fill}}lcd{2.4}d{2.4}d{2.4}cc@{}}
\hline
\textbf{Grid size} & \multicolumn{1}{c}{$\bolds{32\times32}$}
& \multicolumn{1}{c}{$\bolds{64\times64}$} & \multicolumn{1}{c}{$\bolds{128\times128}$}
& \multicolumn{1}{c}{$\bolds{256\times256}$} & \multicolumn{1}{c}{$\bolds{512\times512}$}
& \multicolumn{1}{c@{}}{$\bolds{1024\times1024}$} \\
\hline
$\hat{\theta}^N$ & 1.5355 & 1.5084 & 1.4919 & 1.4975 & 1.5011 & 1.5012\\
& 6.8507 & 6.9974 & 7.1221 & 7.0663 & 6.9841 & 6.9677\\
& 9.2923 & 10.062 & 10.091 & 10.063 & 9.9818 & 9.9600\\
[6pt]
$\sqrt{(\mathcal{I}^{-1})_{ii}}$
& 0.0882 & 0.0406 & 0.0196 & 0.0096 & 0.0048 & 0.0024\\
& 0.5406 & 0.3673 & 0.2371 & 0.1464 & 0.0877 & 0.0512\\
& 0.8515 & 0.5674 & 0.3605 & 0.2202 & 0.1309 & 0.0760\\
[6pt]
$\frac{\sqrt{(\mathcal{G}^{-1})_{ii}}}{\sqrt{(\mathcal{I}^{-1})_{ii}}}$
& 1.0077 & 1.0077 & 1.0077 & 1.0077 & 1.0077 & 1.0077\\
& 1.0062 & 1.0070 & 1.0073 & 1.0074 & 1.0075 & 1.0076\\[2pt]
& 1.0064 & 1.0071 & 1.0073 & 1.0075 & 1.0075 & 1.0076\\
\hline
\end{tabular*}
\end{table}

%f7 #&#

One sees that as the grid becomes larger (denser), the variance of the
estimates decreases as expected. The matrices $\mathcal{I}^{-1}$
and $\mathcal{G}^{-1}$
are comparable in all cases and, in fact, the ratios stay roughly
the same across different sizes of the data. The experiments were run
for data
size up to around one million, and the scaling of the running time
versus data
size is favorable.
The results show a strong agreement of the recorded times with the scaling
$O(n\log n)$.\vfill

%s6 #&#
\section{Application}\label{sec6}
Ozone in the stratosphere blocks ultraviolet radiation from the sun and is
thus essential to all land-based life on Earth.
Satellite-based instruments run by NASA have been measuring total
column ozone
in the atmosphere daily on a near global scale since 1978 (although
with a
significant gap in 1994--1996) and the present instrument is the OMI.
Here, we consider Level 3 gridded data for the month April 2012 in the
latitude band $40^\circ$--$50^\circ$N [Aura OMI Ozone Level-3 Global
Gridded ($1.0\times1.0$ deg) Data Product-OMTO3d (V003)].
Because total column ozone shows persistent patterns of variation with
location, we demeaned the data by, for each pixel,
subtracting off the mean of the available observations during April 2012.
Figure \ref{figozone} displays the resulting demeaned data.
There are potentially $360\times10 = 3600$ observations on each day in this
latitude strip.
However, Figure \ref{figozone} shows 14 or 15 strips of missing
observations each day, which is due to a lack of overlap in OMI observations
between orbits in this latitude band (the orbital frequency of the satellite
is approximately 14.6 orbits per day).
Furthermore, there is nearly a full day of missing observations toward the
end of the record.
For the 30-day period, a complete record would have 108,000
observations, of
which 84,942 are available.

The local time of the Level 2 data on which the Level 3 data are based is
generally near noon due to the sun-synchronous orbit of the satellite, but
there is some variation in local time of Level 2 data
because OMI simultaneously measures
ozone over a swath of roughly 3000 km, so that the actual local times
of the
Level 2 data vary up to about 50 minutes from local noon in the
latitude band
we are considering.
Nevertheless, \citet{fang} showed that, for
Level 3 total column ozone levels (as measured by a predecessor instrument
to the OMI), as long as one stays away from the equator, little
distortion is
caused by assuming all observations are taken at exactly local noon and
we will make this assumption here.
As a consequence, within a given day, time (absolute as opposed to local)
and longitude are completely confounded, which makes distinguishing
longitudinal and temporal dependencies difficult.
Indeed, if one analyzed the data a day at a time, there would be
essentially no
information for distinguishing longitude from time, but by considering
multiple days in a single analysis, it is possible to distinguish their
influences on the dependence structure.

Fitting various Mat\'ern models to subsets of the data within a day, we found
that the local spatial variation in the data is described quite well by the
Whittle model (the Mat\'ern model with smoothness parameter 1) without a
nugget effect.
Results in \citet{stein07} suggest some evidence for spatial anisotropy in
total column ozone at midlatitudes, but the anisotropy is not severe in the
band $40^\circ$--$50^\circ$N and we will ignore it here.
The most striking feature displayed in Figure \ref{figozone} is the obvious
westerly flow of ozone across days.

Based on these considerations, we propose the following simple model for
the demeaned data $Z(\mathbf{x},t)$.
Denoting by $r$ the radius of the Earth, $\varphi$ the latitude, $\psi
$ the
longitude, and $t$ the time, we assume
$Z$ is a 0 mean Gaussian process with covariance function
(parameterized by $\theta_0$, $\theta_1$, $\theta_2$ and $v$):
\[
\cov\bigl\{Z(\mathbf{x}_1, t_1), Z(\mathbf{x}_2,
t_2)\bigr\}= \theta_0\matern_1 \biggl(\sqrt
{\frac{T^2}{\theta_1^2}+\frac
{S^2}{\theta_2^2}} \biggr),
\]
where $T=t_1-t_2$ is the temporal difference,
$S=\|\mathbf{x}(r,\varphi_1, \psi_1-vt_1)-\mathbf{x}(r,\varphi_2,\allowbreak \psi
_2-vt_2)\|$ is the
(adjusted for drift)\vspace*{1pt} spatial difference
and $\mathbf{x}(r,\varphi, \psi)$ maps a spherical coordinate to $\R^3$.
Here, $\matern_{\nu}$ is the Mat\'{e}rn correlation function
%
%e6.1 #&#
\begin{equation}\label{spacetime}
\matern_{\nu}(x)=\frac{(\sqrt{2\nu}x)^{\nu}\besselk_{\nu}(\sqrt{2\nu
}x)}{2^{\nu-1}\Gamma(\nu)}
\end{equation}
with $\besselk_{\nu}$ the modified Bessel function of the second kind of
order $\nu$.
We used the following unit system: $\varphi$ and $\psi$
are in degrees, $t$ is in days, and $r\equiv1$.
In contravention of standard notation, we take longitude to increase as
one heads westward in order to make longitude increase with time within
a day.
Although the use of Euclidean distance in $S$ might be viewed as problematic
[\citet{gneiting}], it is not clear that great circle distances are any more
appropriate in the present circumstance in which there is strong zonal flow.
The model (\ref{spacetime}) has the virtues of simplicity and of validity:
it defines a valid covariance function on the sphere${}\times{}$time whenever
$\theta_0,\theta_1$ and $\theta_2$ are positive.
A more complex model would clearly be needed if one wanted to consider the
process on the entire globe rather than in a narrow latitude band.

Because the covariance matrix $K(\theta_0,\theta_1,\theta_2,v)$ can
be written
as $\theta_0M(\theta_1,\break\theta_2,v)$, where the entries of $M$
are\vspace*{1pt}
generated by
the Mat\'{e}rn function, the estimating equations (\ref{ascore}) give
$\hat\theta_0=Z'M(\hat{\theta}_1,\hat{\theta}_2,\hat
{v})^{-1}Z/n$ as the MLE of $\theta_0$
given values for the other parameters. Therefore, we only need to
solve (\ref{ascore}) with respect to $\theta_1$, $\theta_2$ and~$v$.
Initial values for the parameters were obtained by applying a simplified
fitting procedure to a subset of the data.

We first fit the model using observations from one latitude at a time.
Since there are about 8500 observations per latitude band,
it is possible, although challenging, to compute the exact
MLEs for the observations within a single band using the Cholesky
decomposition.
However, we chose to solve (\ref{ascore}) with the number $N$ of
i.i.d. symmetric Bernoulli vectors $U_j$ fixed at 64. A first order finite
difference filtering [\citet{steinchenanitescufiltering}] was
observed to be
the most effective in encouraging the convergence of the linear solve.
Differences across gaps in the data record were included, so the resulting
sizes of the filtered data sets were just one less than the number of
observations available in each longitude.
Under our model, the covariance matrix of the observations within a
latitude can be embedded in a circulant matrix of dimension 21,600, greatly
speeding up the necessary matrix--vector multiplications.
Table \ref{tabozoneonelat} summarizes the resulting estimates and
the Fisher
information for each latitude band.
The estimates are consistent across latitudes and do not show any obvious
trends with latitude except perhaps at the two most northerly latitudes.
The estimates of $v$ are all near $-7.5^\circ$, which qualitatively
matches the westerly flow seen in Figure~\ref{figozone}.
The differences between $\sqrt{(\mathcal{G}^{-1})_{ii}}/\sqrt{(\mathcal
{I}^{-1})_{ii}}$ and
$1$ were all less than $0.01$, indicating that the choice of $N$ is
sufficient.

%t2 #&#
\begin{table}
\caption{Estimates and standard errors for each latitude}
\label{tabozoneonelat}
\begin{tabular*}{\tablewidth}{@{\extracolsep{\fill}}lccd{2.3}ccccc@{}}
\hline
& & & & & \multicolumn{4}{c@{}}{$\bolds{\sqrt{(\mathcal
{I}^{-1})_{ii}}}$} \\[-4pt]
&&&&& \multicolumn{4}{c@{}}{\hrulefill}\\
\textbf{Latitude} & $\bolds{\hat{\theta}_0^N}$ {$\bolds{(\times}$\textbf{10}$\bolds{^3)}$}
& $\bolds{\hat{\theta}_1^N}$ & \multicolumn{1}{c}{$\bolds{\hat{\theta
}_2^N}$} & $\bolds{\hat{v}^N}$ & {$\bolds{(\times}$\textbf{10}$\bolds{^3)}$} &
& & \\
\hline
$40.5^\circ$N & 1.076 & 2.110 & 11.466 & $-6.991$ & 0.106 &
0.127 & 0.586 & 0.244\\
$41.5^\circ$N & 1.182 & 2.172 & 11.857 & $-6.983$ & 0.123 &
0.136 & 0.634 & 0.251\\
$42.5^\circ$N & 1.320 & 2.219 & 12.437 & $-7.118$ & 0.144 &
0.145 & 0.698 & 0.266\\
$43.5^\circ$N & 1.370 & 2.107 & 12.104 & $-7.369$ & 0.145 &
0.136 & 0.660 & 0.285\\
$44.5^\circ$N & 1.412 & 2.059 & 11.845 & $-7.368$ & 0.145 &
0.130 & 0.628 & 0.294\\
$45.5^\circ$N & 1.416 & 2.010 & 11.814 & $-7.649$ & 0.147 &
0.128 & 0.632 & 0.313\\
$46.5^\circ$N & 1.526 & 2.075 & 12.254 & $-8.045$ & 0.166 &
0.138 & 0.686 & 0.320\\
$47.5^\circ$N & 1.511 & 2.074 & 11.939 & $-7.877$ & 0.161 &
0.135 & 0.654 & 0.319\\
$48.5^\circ$N & 1.325 & 1.887 & 10.134 & $-7.368$ & 0.128 &
0.114 & 0.505 & 0.303\\
$49.5^\circ$N & 1.246 & 1.846 & 9.743 & $-7.120$ & 0.117 &
0.110 & 0.473 & 0.305\\
\hline
\end{tabular*}
\end{table}

The following is an instance of the asymptotic correlation matrix,
obtained by
normalizing each entry of $\mathcal{I}^{-1}$ (at $49.5^\circ$N) with
respect to the diagonal:
\[
\left[ %
\matrix{\hphantom{-}1.0000 & \hphantom{-}0.8830 & \hphantom{-}0.9858 & -0.0080
\cr
\hphantom{-}0.8830 & \hphantom{-}1.0000 & \hphantom{-}0.8767 & -0.0067
\cr
\hphantom{-}0.9858 & \hphantom{-}0.8767 & \hphantom{-}1.0000 & -0.0238
\cr
-0.0080 & -0.0067 & -0.0238 & \hphantom{-}1.0000
}
\right].
\]
We see that $\hat\theta_0,\hat\theta_1$ and $\hat\theta_2$ are all
strongly correlated.
The high correlation of the estimated range parameters $\hat\theta_1$ and
$\hat\theta_2$ with the estimated scale $\hat\theta_0$ is not unexpected
considering the general difficulty of distinguishing scale and range
parameters for strongly correlated spatial data [\citet{zhang}].
The strong correlation of the two range parameters is presumably due to the
near confounding of time and longitude for these data.

Next, we used the data at all latitudes and progressively increased the number
of days.
In this setting, the covariance matrix of the observations can be
embedded in
a block circulant matrix with blocks of size $10\times10$ corresponding
to the 10 latitudes.
Therefore, multiplication of the covariance matrix times a vector can
be accomplished with a discrete Fourier transform for each pair of
latitudes, or ${10 \choose2} = 55$ discrete Fourier transforms.
Because we are using the Whittle covariance function as the basis of our
model, we had hoped filtering the data using the Laplacian would be an
effective preconditioner.
Indeed, it does well at speeding the convergence of the linear solves, but
it unfortunately appears to lose most of the information in the data for
distinguishing spatial from temporal influences, and thus is unsuitable for
these data.
Instead, we used a banded approximate inverse
Cholesky factorization [\citet{kolo}, (2.5), (2.6)]
to precondition the linear solve.
Specifically, we ordered the observations by time and then, since observations
at the same longitude and day are simultaneous, by latitude
south to north.
We then obtained an approximate inverse by subtracting off the conditional
mean of each observation given the previous 20 observations, so the
approximate Cholesky factor has bandwidth 21.
We tried values besides 20 for the number of previous observations on
which to
condition, but 20 seemed to offer about the best combination of fast computing
and effective preconditioning.
The number $N$ of i.i.d. symmetric Bernoulli vectors $U_j$ was increased
to 128, in
order that the differences between
$\sqrt{(\mathcal{G}^{-1})_{ii}}/
\sqrt{(\mathcal{I}^{-1})_{ii}}$ and $1$ were around
$0.1$. The results are summarized in Table \ref{tabozonealllat}.
One sees
that the estimates are reasonably consistent with those shown in
Table \ref{tabozoneonelat}.
Nevertheless, there are some minor discrepancies such as estimates of $v$
that are modestly larger (in magnitude) than found in
Table~\ref{tabozonealllat}, suggesting that taking account of correlations
across latitudes changes what we think about the advection of ozone
from day
to day.

%t3 #&#
\begin{table}
\caption{Estimates and standard errors for all ten latitudes}
\label{tabozonealllat}
\begin{tabular*}{\tablewidth}{@{\extracolsep{\fill}}lcccccccc@{}}
\hline
& & & & & \multicolumn{4}{c@{}}{$\bolds{\sqrt{(\mathcal
{I}^{-1})_{ii}}}$} \\[-4pt]
& & & & & \multicolumn{4}{c@{}}{\hrulefill} \\
\textbf{Days} & $\bolds{\hat{\theta}_0^N}$ {$\bolds{(\times}$\textbf{10}$\bolds{^3)}$}
& $\bolds{\hat{\theta}_1^N}$ & $\bolds{\hat{\theta
}_2^N}$ & $\bolds{\hat{v}^N}$ & {$\bolds{(\times}$\textbf{10}$\bolds{^3)}$} &
& & \\
\hline
\multicolumn{9}{@{}c@{}}{i.i.d. $U_j$'s}\\[4pt]
1--3 & $1.594$ & $2.411$ & $12.159$ & $-8.275$ & $0.362$ & $0.334$ &
$1.398$ & $0.512$\\
1--10 & $1.301$ & $1.719$ & $11.199$ & $-8.368$ & $0.146$ & $0.121$ &
$0.639$ & $0.407$\\
1--20 & $1.138$ & $1.774$ & $10.912$ & $-9.038$ & $0.090$ & $0.085$ &
$0.436$ & $0.252$\\
1--30 & $1.265$ & $1.918$ & $11.554$ & $-8.201$ & $0.089$ & $0.081$ &
$0.414$ & $0.198$\\
[6pt]
\multicolumn{9}{@{}c@{}}{dependent $U_j$'s}\\[4pt]
1--30 & $1.260$ & $1.907$ & $11.531$ & $-8.211$ & $0.088$ & $0.079$ &
$0.406$ & $0.200$\\
\hline
\end{tabular*}
\end{table}

Note that the approximate inverse Cholesky decomposition,
although not as computationally efficient as applying the discrete
Laplacian, is a full rank transformation and thus does not throw out any
statistical information.
The method does require ordering the observations, which is convenient
in the
present case in which there are at most 10 observations per time point.
Nevertheless, we believe this approach may be attractive more generally,
especially for data that are not on a grid.

We also estimated the parameters using the dependent sampling scheme described
in Section \ref{sec3} with $N=128$ and obtained estimates given in the last row of
Table \ref{tabozonealllat}.
It is not as easy to estimate $B^d$ as defined in Theorem \ref{tdependent}
as it is to estimate $B$ with independent $U_j$'s.
We have carried out limited numerical calculations by repeatedly calculating
$g^d(\hat{\theta},N)$ for $\hat{\theta}$ fixed at the estimates
for dependent samples of size $N=128$
and have found that the advantages of using the dependent sampling are
negligible in this case.
We suspect that the reason the gains are not as great as those shown in
Figure~\ref{figN} is due to the substantial correlations of
observations that
are at similar locations a day apart.

%%
%%t4 #&#
%dependent design ($N=16$)}
%%
%Days & $\hat{\theta}_0^N$ & $\hat{\theta}_1^N$ & $\hat{\theta
%}_2^N$ & $\hat{v}^N$ & \multicolumn{4}{c}{$\sqrt{(\mathcal
%{I}^{-1})_{ii}}$} \\
%& {\scriptsize$(\times10^3)$} & & & & {\scriptsize$(\times10^3)$} &
%& & \\
%1--3 & $1.4332$ & $2.3116$ & $11.5190$ & $-8.3703$ & $0.2823$ &
%$0.2863$ & $1.1500$ & $0.5039$\\
%1--10 & $1.3525$ & $1.7453$ & $11.4252$ & $-8.5233$ & $0.1544$ &
%$0.1240$ & $0.6614$ & $0.4077$\\
%1--20 & $1.1548$ & $1.7814$ & $10.9946$ & $-9.0877$ & $0.0933$ &
%$0.0876$ & $0.4509$ & $0.2522$\\
%1--30 & $1.2989$ & $1.9330$ & $11.7133$ & $-8.3067$ & $0.0974$ &
%$0.0855$ & $0.4449$ & $0.2000$\\
%Days & \multicolumn{4}{c|}{$\sqrt{(\mathcal{G}^{-1})_{ii}}$} &
%& {\scriptsize$(\times10^3)$} & & & & {\scriptsize$(\times10^3)$} &
%& & \\
%1--3 & $0.4294$ & $0.4158$ & $1.7521$ & $0.5627$ & $0.4258$ & $0.4183$
%& $1.7362$ & $0.5408$\\
%1--10 & $0.2346$ & $0.1771$ & $1.0069$ & $0.4472$ & $0.2342$ & $0.1785$
%& $1.0045$ & $0.4601$\\
%1--20 & $0.1351$ & $0.1196$ & $0.6539$ & $0.2746$ & $0.1295$ & $0.1154$
%& $0.6262$ & $0.2682$\\
%1--30 & $0.1641$ & $0.1359$ & $0.7510$ & $0.2190$ & $0.1382$ & $0.1137$
%& $0.6324$ & $0.2266$\\
%%
%%

%s7 #&#
\section{Discussion}\label{sec7}
We have demonstrated how derivatives of the
loglikelihood function for a Gaussian process model can be accurately and
efficiently calculated in situations for which direct calculation of the
loglikelihood itself would be much more difficult. Being able to calculate
these derivatives enables us to find solutions to the score equations and
to verify that these solutions are at least local maximizers of the
likelihood. However, if the score equations had multiple solutions, then,
assuming all the solutions could be found, it might not be so easy to
determine which was the global maximizer. Furthermore,
it is not straightforward to obtain likelihood ratio
statistics when only derivatives of the loglikelihood are available.

Perhaps a more critical drawback of having only derivatives of the
loglikelihood occurs when using a Bayesian approach to parameter estimation.
The likelihood needs to be known only up to a multiplicative constant,
so, in principle, knowing the gradient of the loglikelihood throughout
the parameter space is sufficient for calculating the posterior distribution.
However, it is not so clear how one might calculate an approximate posterior
based on just gradient and perhaps Hessian values of the loglikelihood at
some discrete set of parameter values. It is even less clear how one could
implement an MCMC scheme based on just derivatives of the loglikelihood.

Despite this substantial drawback, we consider the development of
likelihood methods
for fitting Gaussian process models that are nearly $O(n)$ in time
and, perhaps more importantly, $O(n)$ in memory, to be essential for expanding
the scope of application of these models.
Calling our approach nearly $O(n)$ in time admittedly glosses over
a number of substantial challenges.
First, we need to have an effective preconditioner for the covariance
matrix $K$.
This allows us to treat $N$, the number of random vectors
in the stochastic trace estimator, as a fixed quantity as $n$ increases
and still obtain estimates that are nearly
as efficient as full maximum likelihood.
The availability of an effective
preconditioner also means that the number of iterations of the
iterative solve can remain bounded as $n$ increases.
We have found that $N=100$ is often sufficient and that the number of
iterations needed for the iterative solver to converge to a tight tolerance
can be several dozen, so writing $O(n)$ can hide a factor of several thousand.
Second, we are assuming that matrix--vector multiplications can be done
in nearly $O(n)$ time.
This is clearly achievable when the number of nonzero entries in $K$
is $O(n)$ or when observations form a partial grid and a
stationary model is assumed so that circulant embedding applies.
For dense, unstructured matrices, fast multipole methods can achieve this
rate, but the method is only approximate and the overhead in the computations
is substantial so that $n$ may need to be very large for the method to be
faster than direct multiplication.
However, even when using exact multiplication, which requires $O(n^2)$
time, despite the need for $N$ iterative solves,
our approach may still be faster than computing the Cholesky
decomposition, which requires $O(n^3)$ computations.
Furthermore, even when $K$ is dense and unstructured,
the iterative algorithm is $O(n)$ in memory, assuming that elements of
$K$ can be calculated as needed, whereas the Cholesky decomposition
requires $O(n^2)$ memory.
Thus, for example, for $n$ in the range 10,000--100,000, even if $K$ has
no exploitable structure, our approach to approximate maximum likelihood
estimation may be much easier to implement on the current generation
of desktop computers than an approach that requires
calculating the Cholesky decomposition of $K$.

The fact that the condition number of $K$ affects both the statistical
efficiency of the stochastic trace approximation and the number of iterations
needed by the iterative solver indicates the importance of having good
preconditioners to make our approach effective. We have suggested a few
possible preconditioners, but it is clear that we have only scratched the
surface of this problem. Statistical problems often yield covariance
matrices with special structures that do not correspond to standard
problems arising in numerical analysis. For example, the ozone data
in Section \ref{sec6} has a partial confounding of time with longitude that made
Laplacian filtering ineffective as a preconditioner. Further development
of preconditioners, especially for unstructured covariance matrices, will
be essential to making our approach broadly effective.

%sA #&#
\begin{appendix}\label{app}
\section*{Appendix: Proofs}
%{\bf Proof}:
%
\begin{pf*}{Proof of Theorem \ref{tmain}}
Since $K$ is positive definite, it can be written in the form $S\Lambda
S'$ with $S$ orthogonal and $\Lambda$ diagonal with
elements $\lambda_1\ge\cdots\ge\lambda_n>0$.
Then $Q^i:= S' K_i S$ is symmetric,
%
%eA.1 #&#
\begin{equation}\label{firstterm}\quad
\operatorname{tr}\bigl(W^i W^j\bigr) = \operatorname{tr}
\bigl(S'K^{-1}SS'K_iSS'K^{-1}SS'K_jS
\bigr) = \operatorname{tr}\bigl(\Lambda^{-1}Q^i\Lambda^{-1}Q^j
\bigr)
\end{equation}
and, similarly,
%
%eA.2 #&#
\begin{equation}\label{secondterm}
\operatorname{tr} \bigl\{W^i \bigl(W^j\bigr)'
\bigr\} = \operatorname{tr}\bigl(\Lambda^{-1}Q^iQ^j
\Lambda^{-1}\bigr).
\end{equation}
For real $v_1,\ldots,v_p$,
%
%eA.3 #&#
\begin{equation}\label{lastterm}
\sum_{i,j=1}^p v_iv_j
\sum_{k=1}^n W_{kk}^iW_{kk}^j
= \sum_{k=1}^n \Biggl\{\sum
_{i=1}^p v_iW_{kk}^i
\Biggr\}^2 \ge0.
\end{equation}
Furthermore, by (\ref{firstterm}),
%
%eA.4 #&#
\begin{equation}\label{firstquad}
\sum_{i,j=1}^p v_iv_j
\operatorname{tr}\bigl(W^i W^j\bigr) = \sum
_{k,\ell=1}^n \frac{1}{\lambda_k\lambda_\ell} \Biggl\{\sum
_{i=1}^p v_iQ^i_{k,\ell}
\Biggr\}^2
\end{equation}
and, by (\ref{secondterm}),
%
%eA.5 #&#
\begin{equation}\label{secondquad}
\sum_{i,j=1}^p v_iv_j
\operatorname{tr} \bigl\{W^i \bigl(W^j\bigr)' \bigr
\} = \sum_{k,\ell=1}^n \frac{1}{\lambda_k^2} \Biggl
\{\sum_{i=1}^p v_iQ^i_{k,\ell}
\Biggr\}^2.
\end{equation}
Write $\gamma_{k\ell}$ for $\sum_{i=1}^p v_iQ^i_{k,\ell}$ and note that
$\gamma_{k\ell}=\gamma_{\ell k}$. Consider finding
an upper bound to
\[
\frac{\sum_{i,j=1}^p v_iv_j\operatorname{tr} \{W^i (W^j)' \}} {
\sum_{i,j=1}^p v_iv_j\operatorname{tr}(W^i W^j)} = \frac{\sum_{k=1}^n
{\gamma_{kk}^2}/{\lambda_k^2} + \sum_{k>\ell}
\gamma_{k\ell}^2 ({1}/{\lambda_k^2} + {1}/{\lambda
_\ell^2} )} {
\sum_{k=1}^n {\gamma_{kk}^2}/{\lambda_k^2} + \sum_{k>\ell}
{2\gamma_{k\ell}^2}/{\lambda_k\lambda_\ell}}.
\]
Think of maximizing this ratio as a function of the $\gamma_{k\ell}^2$'s
for fixed $\lambda_k$'s.
We then have a ratio of two positively
weighted sums of the same positive scalars (the
$\gamma_{k\ell}^2$'s for $k\ge\ell$), so this ratio will be maximized
if the only positive $\gamma_{k\ell}^2$ values correspond to cases for
which the ratio of the weights, here
%
%eA.6 #&#
\begin{equation}\label{ratioweights}
\frac{{1}/{\lambda_k^2}+{1}/{\lambda_\ell^2}}{
{2}/({\lambda_k
\lambda_\ell})} = \frac{1+ ({\lambda_k}/{\lambda_\ell
} )^2} {
{2\lambda_k}/{\lambda_\ell}}
\end{equation}
is maximized.
Since we are considering only $k\ge\ell$, $\frac{\lambda_k}{\lambda
_\ell}
\ge1$ and $\frac{1+x^2}{2x}$ is increasing on $[1,\infty)$, so
(\ref{ratioweights}) is maximized when $k=n$ and $\ell=1$, yielding
\[
\frac{\sum_{i,j=1}^p v_iv_j\operatorname{tr} \{W^i (W^j)' \}} {
\sum_{i,j=1}^p v_iv_j\operatorname{tr}(W^i W^j)} \le\frac{\kappa(K)^2
+1}{2\kappa(K)}.
\]
The theorem follows by
putting this result together with (\ref{B}), (\ref{Jij}) and~(\ref
{lastterm}).
\end{pf*}

\begin{pf*}{Proof of Theorem \ref{tdependent}}
Define $\beta_{ia}$ to be the $a$th element of $\beta_i$
and $X_{\ell a}$ the $a$th diagonal element of $X_\ell$. Then note
that for
$k\ne\ell$ and $k'\ne\ell'$ and $a,b\in\{1,\ldots,N\}$,
\begin{eqnarray*}
& & (U_{i,(k-1)N+a}U_{i,(\ell-1)N+b}, U_{j,(k'-1)N+a'}U_{j,(\ell'-1)N+b'})
\\
& &\qquad = (\beta_{ia}\beta_{ib}Y_{ik}X_{k a}Y_{i\ell}X_{\ell b},
\beta_{ja'}\beta_{jb'}Y_{jk'}X_{k' a'}Y_{j\ell'}X_{\ell'b'})
\end{eqnarray*}
have the same joint distribution as for independent $U_j$'s.
Specifically, the two components are independent symmetric Bernoulli
random variables unless $i=j, a=a', b=b'$ and $k=k'\ne\ell=\ell'$ or
$i=j,a=b',b=a'$ and $k=\ell'\ne\ell=k'$, in which case they are the
same symmetric Bernoulli random variable.
Straightforward calculations yield (\ref{improve}).
\end{pf*}
\end{appendix}

\section*{Acknowledgments}

The data used in this effort were acquired as part of the activities of
NASAs Science Mission Directorate, and are archived and distributed by
the Goddard Earth Sciences (GES) Data and Information Services Center
(DISC).

%suskaldyti doi

% imsref loaded by lrinkeviciute, 2013-04-22 09:53:11
% imsref loaded by lrinkeviciute, 2013-04-22 09:58:39
%

\printaddresses

\end{document}